\documentclass[fleqn,usenatbib]{mnras}

\usepackage{newtxtext,newtxmath}

\usepackage[T1]{fontenc}

\usepackage{graphicx}

\usepackage{multicol}	
\usepackage[english]{babel}
\usepackage{epstopdf}
\usepackage{ae,aecompl}
\usepackage{hyperref}
\usepackage{amsmath}    

\def\mpc{\,h^{-1}{\rm Mpc}}

\def\msun{\,h^{-1}{\rm M}_\odot}
\def\mstar{\,h^{-2}{\rm M}_\odot}

\usepackage{color}

\makeatletter

\newcommand{\Rmnum}[1]{\expandafter\@slowromancap\romannumeral #1@}
\makeatother

\title[galaxy properties around filaments]{Galaxy and halo properties around cosmic filaments
from Sloan Digital Sky Survey Data Release 7 and the ELUCID simulation}

\author[Youcai Zhang]{Youcai Zhang$^{1}$\thanks{yczhang@shao.ac.cn},
Xiaohu Yang$^{2,3}$, Hong Guo$^{1}$, Peng Wang$^{1}$, 
Feng Shi$^{4}$
\\
$^{1}${Shanghai Astronomical Observatory, Nandan Road 80, Shanghai 200030,
  China} \\
$^{2}$State Key Laboratory of Dark Matter Physics, Tsung-Dao Lee Institute \& School of Physics and Astronomy, Shanghai Jiao Tong University, Shanghai 201210, China\\
$^{3}$Shanghai Key Laboratory for Particle Physics and Cosmology, Shanghai 200240, China\\
$^{4}$School of Aerospace Science and Technology, Xidian University, Xi'an 710126, China
}

\begin{document}
\label{firstpage}
\pagerange{\pageref{firstpage}--\pageref{lastpage}}
\maketitle

\begin{abstract}
Using galaxies from the Sloan Digital Sky Survey Data Release 7 (SDSS DR7) along with haloes from the dark matter only constrained ELUCID (Exploring the Local Universe with the reConstructed Initial Density field) simulation, we examine the properties of galaxies and haloes with respect to their distance to cosmic filaments, determined by the medial-axis thinning technique of the COsmic Web Skeleton (COWS) method. Our findings suggest that galaxies or subhaloes grow in mass as they approach these filaments. Galaxies exhibit a redder colour and diminished specific star formation rates as they approach these filaments. Additionally, older subhaloes tend to be more common near the central regions of these filaments. Elliptical galaxies are more frequently found than spiral galaxies in the central regions of the filaments. Lower-mass galaxies typically display reduced sizes in proximity to filaments, whereas higher-mass galaxies tend to exhibit increased sizes when close to filaments. Moreover, the concentration and spin of the haloes grow as they approach the filaments. These findings support the notion that the large-scale structure of the universe, characterized by cosmic web structures, plays a vital role in shaping galaxy and halo properties.

\end{abstract}


\section{Introduction}\label{sec_intro}

In the framework of hierarchical structure formation, galaxies are believed to form within dark-matter haloes,
which reside in a vast and intricate network known as the cosmic web \citep{Bond1996,Cautun2014}. According to the
distribution of galaxies observed in redshift surveys \citep{DESI2024} and dark matter mapped in cosmological
simulations \citep{Pakmor2023}, the cosmic web consists of knots, filaments, sheets, and voids that form due to
gravitational instability from primordial density fluctuations \citep{Peebles1967}. Consequently, the properties
of galaxies and haloes are anticipated to be influenced by the cosmic web environment. The intricate and
fascinating interactions between galaxies, dark-matter haloes, and their surroundings cosmic web present
a complex challenge in understanding galaxy formation and evolution \citep{Wechsler2018, Zhang2021a, Zhang2021b,
Montero2024, WangZitong2024, WangWei2024}.

In recent decades, numerous studies have shown that galaxy properties are influenced not only by their immediate halo
environments but also by the larger cosmic web. On halo scales, galaxies generally exhibit lower star formation rates
(SFRs), redder colours, and more elliptical shapes in dense regions \citep{Woo2013, Old2020}. However, \citet{Taamoli2024}
found that galaxies in dense environments have increased star formation activity at redshifts greater than $z>2$. On larger
scales, galaxies typically experience quenching within cosmic filaments or sheets \citep{Salerno2020, Pasha2023},
although \citet{Hasan2023} suggested that the specific star formation rate (sSFR) does not depend on the cosmic web
environment at $z \geq 2$.

The influence of cosmic web environments on galaxy formation and evolution remains a topic of active discussion \citep{Donnan2022, Rosas2022, Dom2023}. Galaxies are embedded in a complex cosmic web marked by elongated,
thread-like filaments that extend between massive clusters and enclose mostly vacant regions of space known as voids. Many studies have asserted that galaxies located in voids demonstrate higher star formation rates and show bluer colours relative to those in dense environments, with this pattern being especially evident among low-mass galaxies \citep{Hoyle2005, Hoyle2012, Conrado2024, Curtis2024, Parente2024}. However, various research findings indicate that the void environment does not significantly affect galaxy properties \citep{Kreckel2015, Wegner2019}. \citet{Alpaslan2015} analysed environmental catalogues from the Galaxy And Mass Assembly (GAMA) survey \citep{Alpaslan2014} and found that stellar mass, rather than environmental classification, predominantly affects galaxy properties. They found that, when controlling for mass, galaxies display similar properties such as effective radius, morphology, and $u-r$ colour, independent of the large-scale environment. In contrast, this study supports the view that the large-scale environment plays a vital role in shaping galaxy characteristics by analysing the properties of galaxies near cosmic filaments.

Filaments, as the most prominent visual entities in the cosmic web, can act as channels for matter to flow into clusters \citep{Kleiner2017, LuY2024,Rowntree2024}. In recent years, various methodologies have been developed to detect cosmic filaments through geometric or topological analysis of particle distributions, such as DisPerSE \citep{Sousbie2011a,Sousbie2011b}; by using Hessian matrices of density, tidal or velocity shear tensors, such as the NEXUS algorithm \citep{Cautun2013}; and through machine learning techniques, such as the Subspace-Constrained Mean-Shift (SCMS) algorithm \citep{Carron2022}. Generally, these filament detection algorithms are created for different study objectives. 

Employing filament detection methods, numerous studies have explored cosmic filaments to understand their influence on the formation and evolution of galaxies \citep{Darvish2014, Vulcani2019, Sarron2019, Parente2024}. A key area of interest is how the properties of galaxies vary with respect to their distance to the filament axes. Many studies suggest that galaxies often exhibit an increase in stellar mass as they approach filaments based on both simulations \citep{Bulichi2024} and observational data
\citep{ChenYC2017,Malavasi2017,Hoosain2024}. In contrast, \citet{Kuutma2017} found no evidence of an increase in galaxy stellar mass close to filaments, using a filament catalogue generated via the Bisous model point process from SDSS DR10.

Apart from the dependence on stellar mass, galaxies near filaments show redder colours and lower rates of star formation \citep{Tugay2023}, which is confirmed by the filament catalogues generated using the DisPerSE algorithm in various surveys, such as the GAMA survey for $0.02< z < 0.25$ \citep{Kraljic2018}, the Cosmic Evolution Survey (COSMOS) for $0.5 < z < 0.9$ \citep{Laigle2018}, the COSMOS $\ion{H}{I}$ Large Extragalactic Survey (CHILES) for $0 < z < 0.45$ \citep{Luber2019}, the Sloan Digital Sky Survey (SDSS) for $z \sim 0.1$ \citep{Winkel2021}, and the combined data sets from the Wide Field Infrared Survey Explorer (WISE) and SuperCOSMOS for $0.1 < z < 0.3$ \citep{Bonjean2020}. However, \citet{Bulichi2024} found that the quenching of star formation near filaments vanishes, and instead, there is a rise in star formation rates near filaments at $z=2$ for filament catalogues identified with the DisPerSE algorithm in the SIMBA simulation. Based on the IllustrisTNG simulation, \citet{Hasan2024} asserted that the median sSFR relative to the distance to the filament axes depends on the technique used to generate the density field in the DisPerSE algorithm. According to \citet{Hasan2024}, the employment of the Delaunay Tessellation Field Estimator (DTFE) to analyze galaxies reveals a decrease in star formation in galaxies that are proximate to filaments. In contrast, filaments determined by the Monte Carlo Physarum Machine (MCPM) method seem to have little influence on star-forming activities. The observed differences between the DTFE and MCPM methods stem from the way they each identify filamentary structures. The MCPM density field, when used with the DisPerSE method, excels at detecting less prominent filaments that the DTFE density construction method tends to miss.

Galaxy sizes represent a vital observational aspect crucial for enhancing theories on galaxy evolution. \citet{Jiang2019} utilized data from two separate cosmological hydrodynamical simulations, revealing that the size of a galaxy does not depend on the halo spin but inversely correlates with the halo concentration. Considering the substantial influence of cosmic filaments on spin orientation
\citep{Zhang2009, Zhang2013, Zhang2015}, it is intriguing to examine how galaxy and halo sizes, in addition to halo spin and concentration, change relative to their distance from the filament.

This study mainly examines how the properties of galaxies and halos change with respect to their proximity to the filament axes. The galaxy characteristics analyzed are stellar mass, $^{0.1}{g-r}$ color indices, specific star formation rate (sSFR), and half-light radius $R_{50}$, along with the ratio of elliptical to spiral galaxies (E/S ratio). For haloes, this research focuses on the subhalo mass, the redshift of the formation of the subhalo, the half-mass radius of the halo $R_{1/2}$, the virial radius of the halo $R_{\rm vir}$, the spin parameter of the halo $\lambda$ and the concentration of the halo $c_{\rm vir}$. 

One approach to grasp the characteristics of galaxies and halos in relation to filaments is to employ a simulation that is designed based on the real observed density field of galaxy distribution. With the help of the constrained ELUCID simulation, observed galaxies can be efficiently linked to dark matter haloes within the simulation \citep{Yang2018}. The topology and geometry of the cosmic web in simulation are precisely mirrored in comparison to actual observations, as the initial
conditions of the ELUCID simulation are constrained by the group mass density field from SDSS DR7 \citep{Yang2007, Yang2012, WangHuiyuan2012,WangHuiyuan2014, WangHuiyuan2016, WangHY2018}.

The characteristics of filaments are shaped not only by the identification methods used but also by the different tracers, such as dark-matter particles and galaxies \citep{Malavasi2020, Rost2020, Zakharova2023, Zhang2024}.
In this study, we employ dark matter particles from the ELUCID simulation to detect cosmic web filaments using the COWS technique \citep{Pfeifer2022}. Applying a medial axis thinning algorithm \citep{Zhang1984,Lee1994}, cosmic filaments are identified within the cosmic web environment classified by the Hessian matrix of the density field \citep{Zhang2024}. Based on the filament catalogues extracted from the ELUCID simulation, this study investigates the variation in properties of galaxies and haloes relative to their distances to filament axes, using galaxies from SDSS DR7 and haloes in the simulation.

The structure of this paper is organised as follows. Section~\ref{sec_data} describes the observational data from SDSS DR7 and the simulation data from the ELUCID simulation. Additionally, we explain the COWS filament finding technique and the approach used to match observed galaxies with haloes in the constrained simulation. Section~\ref{sec_result} explores how the properties of galaxies and haloes depend on their distance to filaments. Finally, Section~\ref{sec_summary} summarises our findings. In this study, we adopt the cosmological parameters from the ELUCID simulation: $\Omega_m = 0.258$, $\Omega_b = 0.044$, $\Omega_\Lambda = 0.742$, $h = 0.72$, $n_s = 0.963$, and $\sigma_8 = 0.796$.

\section{Data and Method}\label{sec_data}

\subsection{Observational data}
The galaxy dataset used in this research is sourced from the SDSS \citep{York2000,Abazajian2009}, known as one of the most significant astronomical surveys.  \citet{Blanton2005} utilised multiband imaging along with SDSS DR7 spectroscopic data to build the New York University Value-Added Galaxy Catalogue (NYU-VAGC), which features enhancements over the public releases of the SDSS. The NYU-VAGC catalogue is more appropriate for studying large-scale structure statistics because it reduces systematic calibration errors. From the NYU-VAGC, we collected a total of $639~359$ galaxies with redshifts in the range $0.01 \leq z \leq 0.2$, redshift completeness ${\cal C}_z > 0.7$, and apparent magnitudes brighter than $r \leq 17.72$. The criteria for selecting galaxies ensure both reliable $r$-band magnitudes and precise redshift measurements. In addition, the selection criteria ensure that most of the galaxies selected are included in the group catalogue identified via the halo-based group finder from SDSS DR7 \citep{Yang2007, Yang2012}. This catalogue is fundamental for developing the reconstructed initial conditions for the ELUCID simulation \citep{WangHuiyuan2016}.

The galaxy stellar mass $M_*$ is derived using the mass-to-light ratio and color correlation as described by \citet{Bell2003}, which is given by the equation
\begin{equation}\label{eq:stellar_mass}
\begin{split}
\log \left( \frac{M_*}{h^{-2} M_\odot} \right) =& -0.306 + 1.097 [^{0.0}(g-r)] 
    - 0.1 \\& - 0.4(^{0.0}M_r - 5 \log h - 4.64),
\end{split}
\end{equation}
where $^{0.0}(g-r)$ and $^{0.0}M_r - 5 \log h$ represent the $(g-r)$ color and $r$-band magnitude corrected for $K+E$ at $z=0.0$, with $4.64$ being the $r$-band magnitude of the Sun in the AB magnitude system, and the $-0.10$ factor indicating the use of a \citet{Kroupa2001} initial mass function.

The galaxy star formation rates are sourced from the public catalogue of \citet{Chang2015},
which includes reliable aperture corrections. 
They combine SDSS and Wide-field Infrared Survey Explorer \citep[WISE;][]{Wright2010} photometry for the entire SDSS spectroscopic galaxy sample to construct spectral energy distributions (SEDs). \citet{Chang2015} employ the Multi-wavelength Analysis of Galaxy Physical Properties
\citep[MAGPHYS\footnote{\href{https://www.iap.fr/magphys/}{\tt https://www.iap.fr/magphys/}};][]{daCunha2008} technique to coherently and systematically model both the attenuated stellar SED and the dust emission. This approach enables more reliable estimations of galaxy star formation rates by considering emissions from polycyclic aromatic hydrocarbons (PAH) as well as thermal dust radiation. The stellar masses listed in the public catalogue are also used to explore the relationship between the mass of galaxies and their proximity to filament spines. Overall, the principal conclusions remained stable despite the variations in the stellar mass applied.

The morphological classifications of galaxies originate from the Galaxy Zoo 2 catalogue \citep[GZ2\footnote{\href{https://data.galaxyzoo.org/}{\tt https://data.galaxyzoo.org/}};][]{Willett2013}, which uses classifications from volunteer citizen scientists to measure morphologies
for galaxies in the SDSS DR7 Legacy Survey with apparent magnitudes $r\leq17$.  
A multilevel decision tree was used to classify galaxy morphologies, drawing on inputs from citizen scientist classifiers through an online platform. The comprehensive catalogue includes morphological classifications for $304~122$ galaxies, each galaxy being associated with strings identified as $gz2\_class$.
Galaxies classified as elliptical and spiral are represented by $gz2\_class$ strings that start with 'E' and 'S', respectively. 
The data set generated by matching the GZ2 catalogue with the NYU-VAGC catalogue comprises $107~230$ elliptical galaxies and $134~024$ spiral galaxies.

Galaxy sizes are obtained from the NYU-VAGC catalogue within the table $\it{object\_sdss\_imaging.fits}$, focussing on the $r$-band radii denoted as $R_{50}$, which encompass $50\%$ of the Petrosian flux. The Petrosian flux $F_{\rm P}$ for any band is the flux within a certain number $N_{\rm P}$ ($N_{\rm P}=2.0$ in the SDSS) of the Petrosian radius $r_{\rm P}$, and is given by:
\begin{equation}\label{petro_flux}
F_{\rm P} = \int_0^{N_{\rm P} r_{\rm P}} 2\pi r I(r) {\rm d}r,
\end{equation}
where $I(r)$ is the average surface brightness in relation to radius, and $r_{\rm P}$ is the radius at which the Petrosian ratio $R_{\rm P}$ reaches a specified limit ($R_{\rm P} = 0.2$ in the SDSS). The ratio $R_{\rm P}$ is defined as the ratio of
the local surface brightness in an annulus to the mean surface brightness within the radius $r_{\rm P}$, expressed by \citep{Blanton2001}:
\begin{equation}\label{petro_radius}
R_{\rm P}= \frac {\int_{0.8r_{\rm P}}^{1.25r_{\rm P}} 2\pi r I(r) {\rm d}r /[\pi(1.25^2-0.8^2)r_{\rm P}^2] } 
{\int_{0}^{r_{\rm P}} 2\pi r I(r) {\rm d}r /(\pi r_{\rm P}^2) }.
\end{equation}
In general, the galaxy sizes listed in the
NYU-VAGC catalogue agree well with those reported in the public catalogues by \citet{Simard2011}
and \citet{Meert2015}, particularly for galaxies with lower masses\citep{Zhang2022}. Unless
mentioned otherwise, the galaxy sizes from the NYU-VAGC catalogue are used throughout the paper.

\subsection{Simulation data}

The ELUCID simulation, which focusses only on dark matter, is a constrained simulation capable of replicating the density field of the nearby universe \citep{WangHuiyuan2012, WangHuiyuan2014, WangHuiyuan2016}. These reconstructions are constrained by the distribution of dark-matter haloes, which are represented by galaxy groups exceeding a mass of $10^{12}\msun$, based on data from the SDSS redshift survey \citep{Yang2007,Yang2012}.  

The initial conditions for the density field stem from reconstruction techniques that utilise the halo domain approach and the Monte Carlo Hamiltonian Markov chain method (HMC) \citep{WangHuiyuan2012, WangHuiyuan2014}. In the context of the HMC framework, Fourier modes are specifically created for smaller $k$ values, which correspond to larger scales, with a particular focus when $k \leq \pi/l_g$, where $l_g$ denotes the grid cell size used in the HMC process. Random phases are used to compensate for the absence of higher $k$ modes, which represent smaller scales. Therefore, about $95\%$ of the galaxy groups with masses exceeding $10^{14}\msun$ can be accurately aligned with simulated halos of comparable masses, allowing a distance error of up to $4\mpc$ \citep{WangHuiyuan2016,Chen2019}. Massive structures such as the Coma cluster and the Sloan Great Wall are effectively replicated in this reconstruction \citep{Luo2024}. However, caution must be exercised when employing small-scale structures from the ELUCID simulation to depict observed galaxy distributions \citep{Tweed2017}.

The simulation begins with an initial condition sampled using $3072^3$ dark-matter particles at a
redshift of $z = 100$ within a periodic box measuring $500 \mpc$ per side. The density field
subsequently evolved to the current epoch using a memory-efficient version of GADGET2
\citep{Springel2005}. Throughout the simulation, $100$ snapshots are generated, ranging from
redshift $z = 18.4$ to $z = 0$, with the expansion factor distributed in logarithmic intervals.
Each dark matter particle has a mass of $3.1 \times 10^8 \msun$.

In the simulation, the conventional Friends-of-Friends (FOF) algorithm \citep{Davis1985} is applied
to the particle data to create FOF haloes with a linking length of $b = 0.2$ times the average
particle separation. Subsequently, the SUBFIND algorithm \citep{Springel2001} breaks down a given
FOF halo into a collection of separate subhaloes with gravitational bind during the unbinding
process, where the gravitational potentials are computed iteratively to eliminate the unbound
particles within the host FOF halo. By identifying subhaloes with the SUBFIND algorithm, merger
trees are built by connecting a subhalo in one snapshot to its descendant subhalo in the next
snapshot. The formation redshift $z_{\rm f}$ of the subhalo is defined as the redshift when the
progenitor of the main branch reached half of its maximum mass $M_{\rm peak}$ \citep{Zhang2021b}.

For each subhalo in the simulation, the dimensionless spin parameter $\lambda$ is
determined using the equation \citep{Peebles1969}
\begin{equation}\label{eqn:spin}
\lambda = \frac { J |E|^{1/2} }{G M_{\rm h}^{5/2}},   
\end{equation}
where $M_{\rm h}$ refers to the subhalo mass, $J$ signifies the total angular momentum, 
$E$ indicates the subhalo's total energy, and $G$ is Newton's gravitational constant.

For a halo characterized by a virial radius $R_{\rm vir}$ \citep{Bryan1998}, the 
concentration is defined as
\begin{equation}\label{eqn:cvir}
c_{\rm vir} = \frac{R_{\rm vir}}{r_{\rm s}},
\end{equation}
where $r_{\rm s}$ denotes the scale radius in the density profiles, as
outlined by the straightforward formula in \citet{Navarro1997}
\begin{equation}\label{eqn:nfw}
\frac {\rho(r)}{\rho_{\rm cirt}} = \frac {\delta_{\rm c}}{(r/r_{\rm s})(1 + r/r_{\rm s})^2},
\end{equation}
where $\rho_{\rm cirt}$ is the critical density of the universe, and 
$\delta_{\rm c}$ represents a characteristic density contrast
\citep{Zhao2009, Zhang2022}.

\subsection{Algorithm for detecting filaments}\label{sec_algorithm}

\begin{figure}
\includegraphics[width=0.5\textwidth]{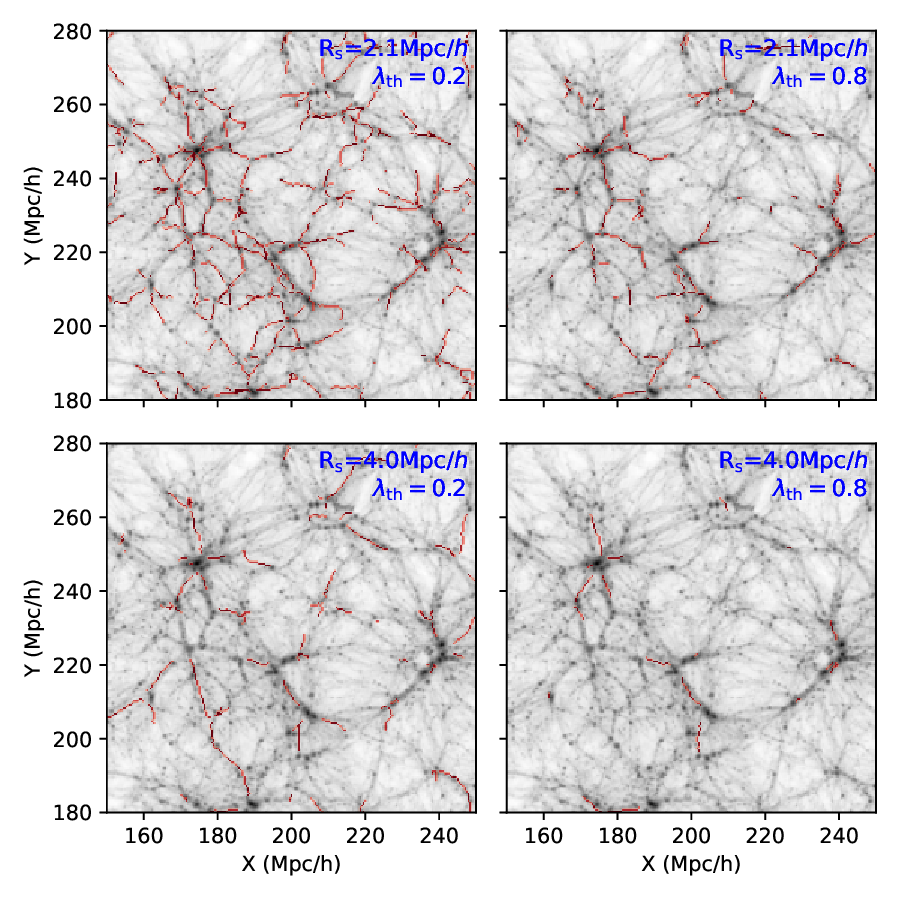}
\caption{ Distribution of density (black) and filament axes (red) within a $10 \mpc$ wide slice from the ELUCID simulation. The COWS technique is employed to identify the filament axes, utilizing cosmic web classifications based on varying values of $R_{\rm s}$ and $\lambda_{\rm th}$. }
\label{fig:slice}
\end{figure}

\begin{figure}
\includegraphics[width=0.5\textwidth]{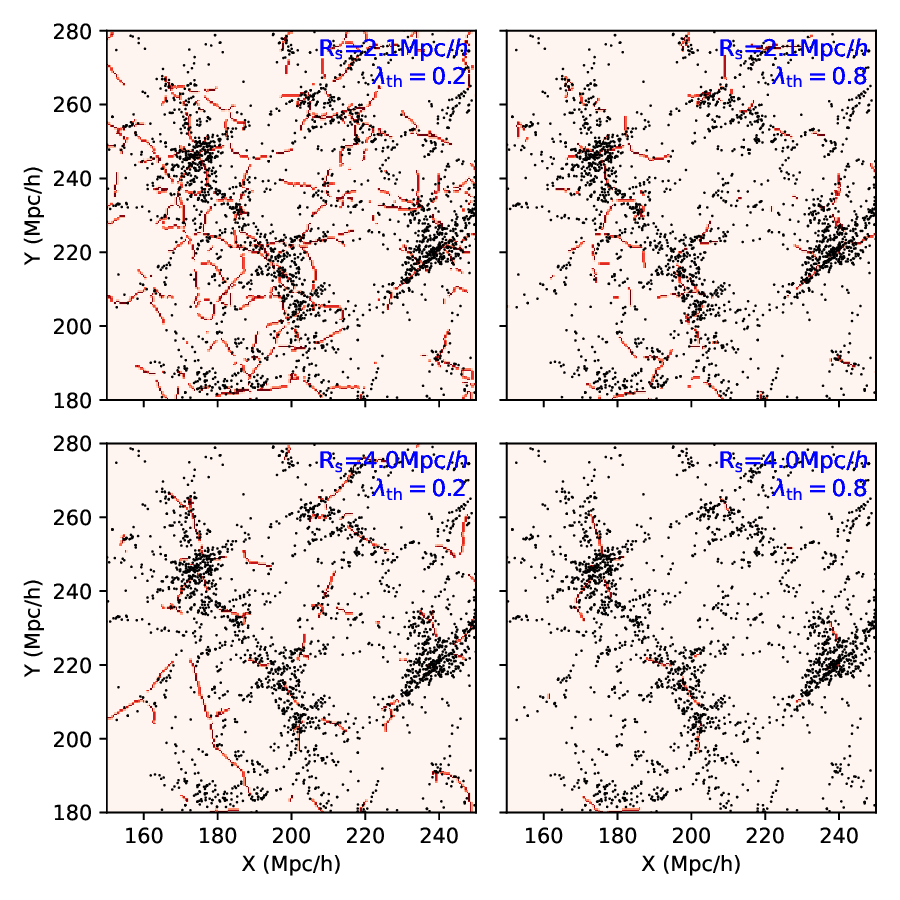}
\caption{ Similar to Figure~\ref{fig:slice}, but for the COWS filaments overlaid on the galaxy distribution from SDSS DR7.}
\label{fig:slice_sdss}
\end{figure}

The spatial environment at any point is classified into knots, filaments, sheets, and voids
employing the Hessian matrix of the density field \citep{Zhang2009}. To build the density
field, a cloud-in-cell (CIC) interpolation with $1024^3$ cells is applied to $3072^3$
dark matter particles in the ELUCID simulation. Subsequently, this field is smoothed using
a Gaussian kernel with a width of $R_{\rm s}$ to produce the smoothed density
$\rho_{\rm s}$. The Hessian matrix of the smoothed density field, normalised
negatively\footnote{The negative sign on the right-hand side of Equation~\ref{eqn:hessian}
is included to guarantee that positive (negative) eigenvalues correspond to collapsing
(expanding) matter. This is consistent with the eigenvalue signs of the tidal tensor in
the dynamic model \citep{Hahn2007}.}, is defined for each cell of the grid as
\begin{equation}\label{eqn:hessian}
H_{ij} = - \frac{R_{\rm s}^2}{\rho_{\rm mean}} \frac {\partial^2 \rho_{\rm s}}
{\partial x_i \partial x_j }, 
\end{equation}
where $\rho_{\rm mean}$ is the average density of the universe. 
In Equation~\ref{eqn:hessian}, the eigenvalues of the Hessian matrix at the location of each cell are represented by $\lambda_1, \lambda_2$, and $\lambda_3$ (ordered such that $\lambda_1 > \lambda_2 > \lambda_3$) with corresponding eigenvectors $\hat{\boldsymbol{e}}_1, \hat{\boldsymbol{e}}_2$, and $\hat{\boldsymbol{e}}_3$. The classification of the cosmic web type for each cell is determined by counting how many eigenvalues satisfy $\lambda_i > \lambda_{\rm th}$, where $i$ equals 1, 2, or 3.
The classification of the environment
is specific to each cell, and the grouping of adjacent cells sharing the same cosmic web-type
forms the geometric structures known as voids, sheets, filaments, or knots.

Taking into account the spatial arrangement of the filament cells, the filament axes are
determined using the medial axis thinning algorithm implemented in the COWS method
\citep{Pfeifer2022}. According to cosmic web classifications, the initial data set is converted into a binary format where knot and filament cells are given a value of $1$, while sheets and void cells receive a value of $0$. Subsequently, the medial axis thinning algorithm \citep{Lee1994} is applied to condense the extensive binary data into a single-cell skeleton by detecting and removing border cells, while maintaining the geometric and topological characteristics of the data set. Finally, individual filaments are identified by removing extraneous contaminants or features, such as junctions, hollow cavities, and knot cells, from the topological skeleton.

In previous research \citep{Zhang2024}, we explored the statistical characteristics of filaments using different parameters $R_{\rm s}$ and $\lambda_{\rm th}$.
According to \citet{Zhang2024}, as the smoothing parameter $R_{\rm s}$ increases, the filament lengths tend to be longer. Moreover, our analysis of the density profiles showed that the filaments generally thicken with increasing $R_{\rm s}$.

Figure~\ref{fig:slice} illustrates the distribution of dark matter density and filaments in a $10 \mpc$ slice of the ELUCID simulation, where darker regions represent areas of higher density. The filament spine is detected through the COWS technique, which is based on cosmic web classifications that differ in values $R_{\rm s}$ and $\lambda_{\rm th}$. Figure~\ref{fig:slice} compares filament axes determined with different parameters: the top panels illustrate these axes using a smoothing length of $R_{\rm s}=2.1\mpc$ with thresholds $\lambda_{\rm th}=0.2$ and $0.8$, while the bottom panels present results for $R_{\rm s}=4.0\mpc$. Generally, the filament axes are closely aligned with most of the visible filamentary structures in the darker regions.

To determine the corresponding positions in the ELUCID simulation that align with the observed locations of galaxies, the galaxies' observable coordinates ($\alpha$, $\delta$, z) are transformed into the simulation's (X, Y, Z) coordinate system using the following equations:
\begin{equation}\label{coordinate_trans}
\begin{split}
X & = R(z) \cos \delta \cos \alpha \\
Y & = R(z) \cos \delta \sin \alpha \\
Z & = R(z) \sin \delta,
\end{split}
\end{equation}
where $\alpha$ is the right ascension, $\delta$ is the declination, and $R(z)$ is the comoving distance corresponding to redshift $z$. Figure~\ref{fig:slice_sdss}, similar to Figure~\ref{fig:slice}, shows the COWS filaments overlaid on the galaxy distribution. Here, the observed galaxy positions have been transformed into the coordinate framework of the ELUCID simulation. Typically, the axes of the filaments are nearly parallel to the filamentary structures outlined by the distribution of galaxies.

Consider a defined neighbourhood around a specific cell, such as the 26 neighbourhoods $N_{26}$, which includes all adjacent cells within a $3 \times 3 \times 3$ cube, covering all direct, diagonal and corner neighbours.
\citet{Lee1994} introduced a technique to identify the cells that make up the skeleton by counting the neighbours that each cell possesses in $N_{26}$. Cells with 1, 2, 3, and 4 neighbours are classified as end points, regular cells, T junctions, and X junctions, respectively. For the purpose of skeleton classification, each cell is assigned a number corresponding to the count of neighbouring cells in $N_{26}$. Cells possessing more than two neighbours are removed. Filaments can be detected by recursively connecting cells within $N_{26}$, beginning from a cell marked as an endpoint and continuing until another endpoint is encountered, which means that no further cells can be linked.
Observe that each filament detected by the COWS method has two endpoints linked by a single path with a width of one cell. This configuration allows for the calculation of how galaxy properties vary with the distance perpendicular to the filament axes.

Clearly, the medial axis thinning algorithm used to detect filaments in cosmic web catalogues operates without any free parameters. However, the cosmic web classification technique involves two parameters: the smoothing length $R_{\rm s}$ and the threshold $\lambda_{\rm th}$, both of which can influence the filament catalogue derived from the dark-matter particle density distribution \citep{Zhang2024}. In the following analysis, we focus primarily on the findings for $\lambda_{\rm th}=0.2$ and $0.8$, using the smoothing lengths of $R_{\rm s} = 2.1\mpc$ and $4.0\mpc$. 

Figure~\ref{fig:slice} demonstrates that by increasing $R_{\rm s}$ or $\lambda_{\rm th}$, the filament detection method effectively eliminates a large number of less significant filaments \citep[see also Figures $1$ and $2$ in][]{Zhang2024}. In the filament catalogue, for $R_{\rm s} = 2.1 \mpc$ and $\lambda_{\rm th} = 0.2$ in the SDSS area, there are $390~001$ filament segments. In comparison, for $R_{\rm s} = 4.0 \mpc$ and $\lambda_{\rm th} = 0.8$, the catalogue records $27~978$ segments. 
An increase in $R_{\rm s}$ or $\lambda_{\rm th}$ results in a reduction in filaments.
It is assumed that the spines of the filaments run through the centres of the filament cells. A filament consists of several segments, each segment connecting the centres of adjacent cells along the filament. This research examines the properties of galaxies/haloes around the segments that make up the filament.
 
It is well established that the eigenvectors $\hat{\boldsymbol{e}}_3$ serve as indicators for the directions of the filaments in numerous studies \citep{Zhang2009, Zhang2013, Zhang2015}. According to \citet{Zhang2024}, we confirmed that the filament axes identified by the COWS method align well with the eigenvector $\hat{\boldsymbol{e}}_3$ of the Hessian matrix of the smoothed density field, regardless of the various parameters $R_{\rm s}$ and $\lambda_{\rm th}$. This validates the effectiveness of the COWS method. \citet{Zhang2024} observed that the probability distribution function (PDF) for the filament lengths relies solely on the smoothing length $R_{\rm s}$ and remains largely consistent across various $\lambda_{\rm th}$ thresholds. Furthermore, this study reveals that the characteristics analysed concerning their distance from the filament spines show similar increasing or decreasing trends across various parameters $R_{\rm s}$ and $\lambda_{\rm th}$.
However, the observed trend of decreasing or increasing is relatively weak in the catalogue of filaments characterised by $R_{\rm s} = 2.1\mpc$ and $\lambda_{\rm th}=0.2$.

\section{Results}\label{sec_result}

This section explores the variation in galaxy and halo properties in relation to their distance to filament axes, determined using the medial-axis thinning algorithm as employed in the COWS technique. The attributes of galaxies under investigation include stellar mass, colour, specific star formation rate (sSFR), size, and the elliptical-to-spiral ratio (E/S ratio). For haloes, the characteristics evaluated encompass mass, age, radius, spin, and concentration.

\subsection{Galaxies properties in observation}\label{sec_galaxy}
\subsubsection{Stellar mass}\label{sec_mass}
\begin{figure}
\includegraphics[width=0.5\textwidth]{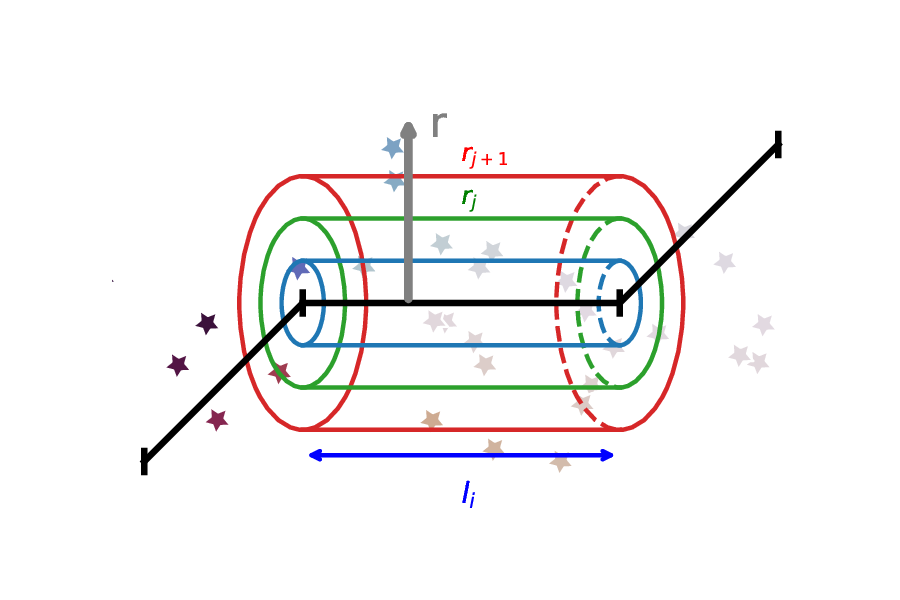}
\caption{
 Illustration of segment $i$ with length $l_i$, useful for determining averaged values in the $j$-th shell, characterised by a thickness of $r_{j+1}-r_{j}$. The radial distance from the axis of the filament is represented by $r$. In this study, each cylindrical segment $i$ is divided into $10$ cylindrical shells.
}
\label{fig:cylinder}
\end{figure}

\begin{figure}
\includegraphics[width=0.5\textwidth]{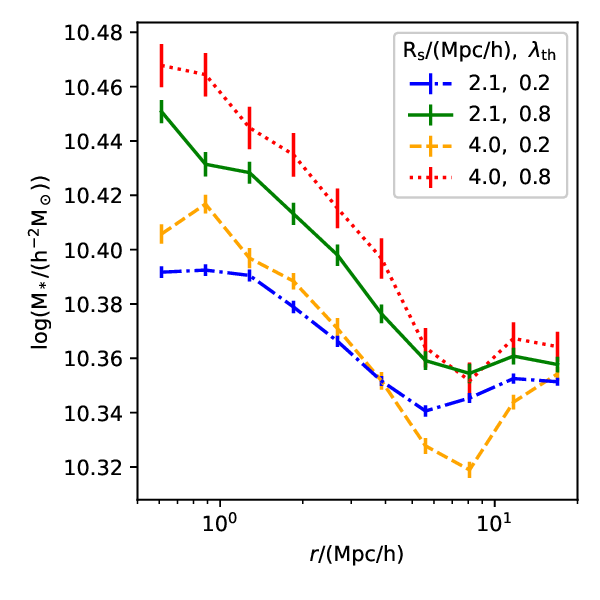}
\caption{Stellar mass of galaxies from SDSS DR7 as a function of their distance to the filament spines identified by the COWS method, using varying parameters for $R_{\rm s}$ and $\lambda_{\rm th}$. The average values of stellar mass are calculated in shells around filaments, illustrated by the different line types, indicating the 
average stellar mass decreases with the distance to filament axes.
The error bars are derived from $100$ bootstrap samples.}
\label{fig:stellar_mass}
\end{figure}

The filament catalogues used in this research are derived from the distribution of dark matter particles within the ELUCID simulation. This simulation replicates the cosmic web structure found in SDSS, as its initial conditions are based on the density field observed in the SDSS data \citep{WangHuiyuan2014}. We assessed the stellar mass close to the axes of filaments as determined in the ELUCID simulation, using the distribution of galaxies, with their observed positions mapped into the ELUCID simulation's coordinate system.

In the algorithm for detecting filaments, the cosmic web environment is categorised for $1024^3$ cells within a $500 \mpc$ box. The filament axes are considered to traverse the centre of each filament cell. However, their precise locations within a $0.5 \mpc$ cell remain uncertain. Therefore, to exclude the core region where the positions of the filament axes are ambiguous, this study sets the minimum profile radius at $0.5 \mpc$. The filament catalogue produced using the COWS method defines a filament as being composed of multiple linked segments. In our study, we analyse the distribution of stellar mass around each segment of these filaments. Each cylindrical segment is divided into $10$ cylindrical shells, employing uniform logarithmic bins centred along the axis. These bins extend from an inner radius of $r=0.5\mpc$ to $r=20\mpc$ for each segment. We then calculated the average stellar mass values of galaxies within each radial shell for all segments listed in the catalogue. A filament is made up of several interconnected segments. As illustrated in Figure~\ref{fig:cylinder}, the average stellar mass for the $j$-th cylindrical shell, with a thickness of $r_{j+1}-r_{j}$, is determined by summing the stellar masses of all galaxies within this shell and dividing by the total count of galaxies across all segments.

Figure~\ref{fig:stellar_mass} illustrates the stellar mass of galaxies from SDSS DR7 in relation to their distance to the filament spines, identified via the COWS method. The different lines represent the average values for the respective scenarios $R_s$ and $\lambda_{\rm th}$. The error bars in Figure~\ref{fig:stellar_mass} are obtained by bootstrapping segment profiles using the \texttt{astropy.stats.bootstrap} function from the \texttt{Astropy} library\footnote{\href{http://www.astropy.org}{http://www.astropy.org}}. A random subset of segments is repeatedly selected from the complete set, and the averages of these selections are calculated. The random sampling and averaging procedure is carried out $100$ times to generate $100$ mean estimates. From these means, the standard error is calculated to create the error bars shown in Figure~\ref{fig:stellar_mass}.

Overall, as illustrated in Figure~\ref{fig:stellar_mass}, galaxies tend to show a growth in stellar mass as they approach filaments, which is consistent with results from previous observations \citep{ChenYC2017,Malavasi2017,Hoosain2024}.
As described in Section~\ref{sec_algorithm}, the approach for determining filaments involves two key parameters: the smoothing length, $R_{\rm s}$, and the threshold, $\lambda_{\rm th}$, both influencing the filament catalogue derived from the dark-matter particle distribution. It is worth exploring how galaxy properties near filaments are affected by these parameters. Figure~\ref{fig:stellar_mass} illustrates how stellar masses are distributed among galaxies close to the filament under four conditions: $R_s=2.1,~4.0\mpc$, and $\lambda_{\rm th}=0.2,~0.8$.

Figure~\ref{fig:stellar_mass} illustrates that with higher values of $R_{\rm s}$ and $\lambda_{\rm th}$, the increase in stellar mass is more pronounced as galaxies approach filaments. This pattern is especially noticeable in the filament catalogue when using the parameters $R_{\rm s}=4.0 \mpc$ and $\lambda_{\rm th}=0.8$, as this configuration retains only the most significant filaments. In the previous study by \citet{Zhang2024}, we analysed the statistical properties of the filaments using different parameters $R_{\rm s}$ and $\lambda_{\rm th}$. We confirmed that the filaments increase both in length and in thickness as $R_{\rm s}$ increases. As the smoothing length $R_{\rm s}$ increases, less prominent filaments are typically omitted from the filament catalogue, which enhances the observed trend of increasing mass closer to filaments for $R_{\rm s}=4.0 \mpc$ and $\lambda_{\rm th}=0.8$.

\subsubsection{Galaxy colour}

\begin{figure}
\includegraphics[width=0.5\textwidth]{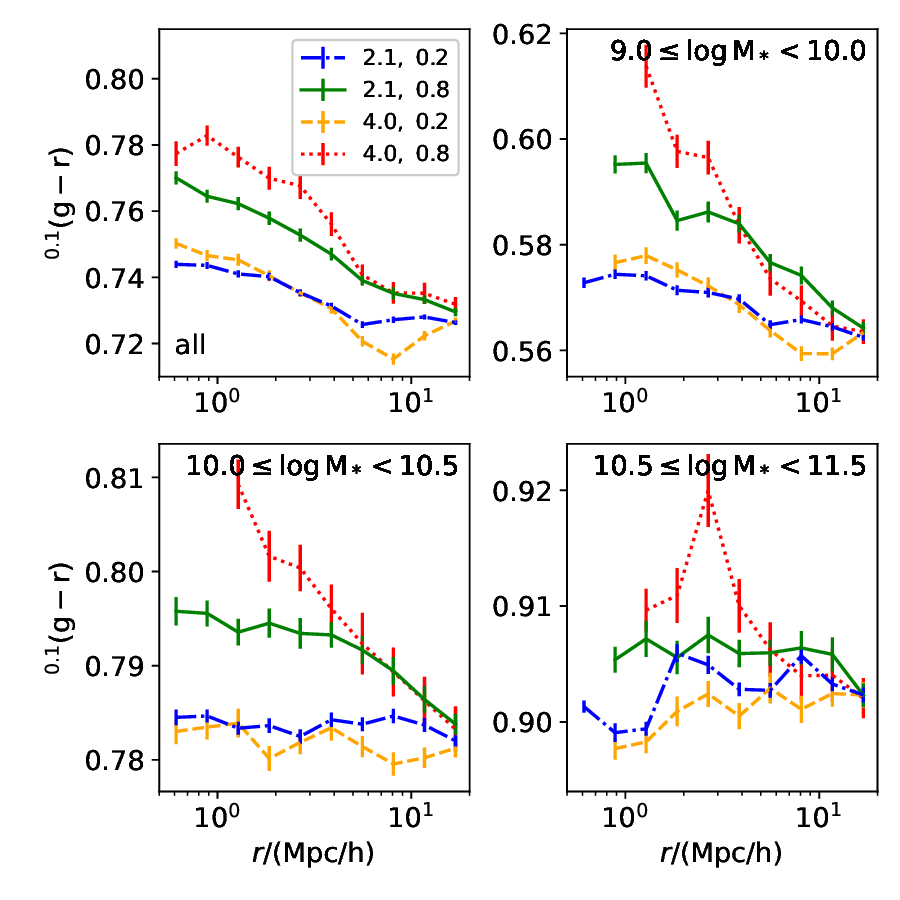}
\caption{Galaxy colors as a function of the distance to the filament spines identified by the COWS method, employing different parameters for $R_{\rm s}=2.1, 4.0\mpc$ and $\lambda_{\rm th}=0.2, 0.8$. The top-left panel presents the findings for all galaxies in the sample, while the other panels depict results divided by galaxy mass bins with $\log (M_*/ (\mstar))$ spanning $(9,10)$, $(10,10.5)$, and $(10.5,11.5)$, respectively.}
\label{fig:colour}
\end{figure}

The colour of a galaxy offer valuable information about its star formation rate, quenching mechanisms, and the age of its stellar population. In the following analysis, we examine how changes in galaxy colour relate to different distances from the centres of filaments.

As mentioned in section~\ref{sec_mass}, the mass of the galaxy is significantly influenced by its proximity to filament axes. To reduce the influence of stellar mass on our findings, we classify galaxies into three separate mass categories specified by $\log (M_*/ (\mstar))$: $(9,10)$, $(10,10.5)$, and $(10.5,11.5)$. These categories contain $129~651$, $195~540$, and $176~551$ galaxies, respectively.

Figure~\ref{fig:colour} illustrates how galaxy colours $^{0.1}(g-r)$ vary with the distance to the filament spines determined using the COWS method, with parameters set at $R_s=2.1, 4.0\mpc$ and $\lambda_{\rm th}=0.2, 0.8$. In Figure~\ref{fig:colour}, the top left panel illustrates the results for the entire galaxy sample with a mass of $\log (M_*/ (\mstar)) \geq 9.0$. The remaining panels show results segmented by galaxy mass bins with $\log (M_*/ (\mstar))$ intervals of $(9,10)$, $(10,10.5)$, and $(10.5,11.5)$, respectively. Overall, galaxies exhibit a redder colour as they approach the cores of filaments, consistent with earlier research \citep{Kraljic2018, Tugay2023}. Furthermore, the change in colour with respect to the distance to the filament spines is more evident with larger $R_{\rm s}$ or $\lambda_{\rm th}$. 

As depicted in Figure~\ref{fig:colour}, within a particular range of stellar masses, galaxies are also redder when they are located close to filaments, especially at $R_{\rm s}=4.0 \mpc$ and $\lambda_{\rm th}=0.8$. However, the correlation between galaxy colours and their distance to filament axes is quite weak for $R_{\rm s}=2.1 \mpc$ and $\lambda_{\rm th}=0.2$.

Galaxies in dense filament environments experience gas stripping, tidal interactions, and quenching mechanisms (e.g., ram-pressure stripping, strangulation), which suppress star formation and lead to older, redder stellar populations. These processes are most efficient in the high-density cores of well-defined filaments \citep{Bulichi2024}.
A more pronounced correlation between galaxy colour and proximity to filament spines at higher $R_{\rm s}$ or $\lambda_{\rm th}$ probably arises from improved detection of prominent filaments. An increased $R_{\rm s}$ leads to a smoothing of the cosmic density field over more extensive scales, thereby averaging out the noise on smaller scales and more effectively revealing the actual geometry of the filaments. For example, at $R_{\rm s}=4.0\mpc$, noise-induced "fake" filaments in low-density regions are smoothed away, leaving only robust filaments. In well-defined filaments, processes like ram-pressure stripping and tidal interactions efficiently suppress star formation, leading to redder galaxies near filaments.

\subsubsection{Specific star formation rate}

\begin{figure}
\includegraphics[width=0.5\textwidth]{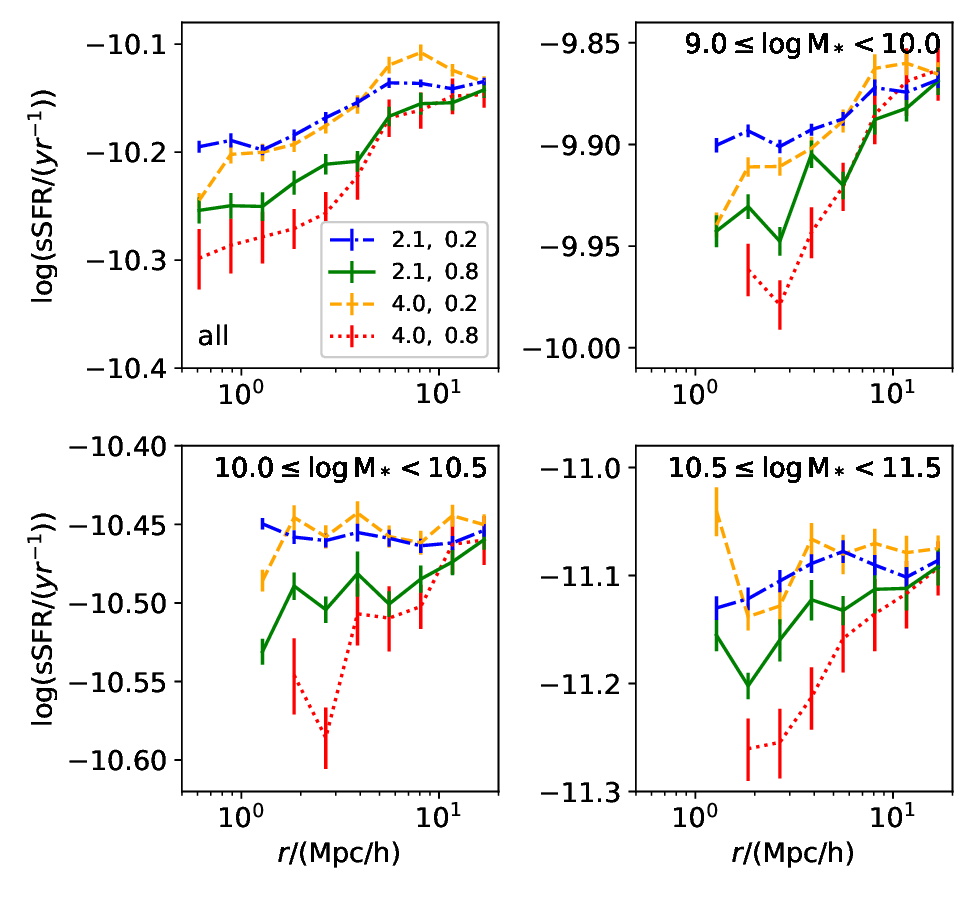}
\caption{Specific star formation rate as a function of the distance to the filament spines identified by the COWS method, using varying parameters for $R_{\rm s}=2.1, 4.0\mpc$ and $\lambda_{\rm th}=0.2, 0.8$. The top-left panel shows the results for the entire galaxy sample, whereas the remaining panels display findings segmented by galaxy mass categories with $\log (M_*/ (\mstar))$ ranges of $(9,10)$, $(10,10.5)$, and $(10.5,11.5)$, respectively.}
\label{fig:ssfr}
\end{figure}

As indicated by \citet{Winkel2021}, a study using the SDSS survey demonstrated that galaxies show increased specific star formation rates (sSFRs) as their separation from cosmic web filaments grows, using a filament catalogue created with the DisPerSE approach. In this research, we investigate how the specific star formation rate (sSFR) varies in relation to the separation from the filaments, employing a catalogue formulated using the COWS methodology. The specific star formation rates for galaxies in our analysis are sourced from the publicly accessible catalogue by \citet{Chang2015},  which offers reliable aperture corrections.

Figure~\ref{fig:ssfr} shows how the specific star formation rate varies with the distance to the filament spines, as determined by the COWS approach. The upper left panel illustrates an increase in the sSFR of the whole sample as the distance from the filament spines grows. This aligns with the results reported in the filament catalogues obtained using the DisPerSE method \citep{Laigle2018, Luber2019, Bonjean2020, Winkel2021}. As shown by the green line at $R_{\rm s}=2.1 \mpc$ and $\lambda_{\rm th}=0.8$ in Figure~\ref{fig:ssfr}, a consistent trend is evident across specified mass intervals $\log (M_*/ (\mstar))$, including $(9,10)$, $(10,10.5)$ and $(10.5,11.5)$. Moreover, the variation in sSFR in relation to the proximity to filament spines becomes more pronounced with higher $R_{\rm s}$ or $\lambda_{\rm th}$ values. This trend is particularly significant for $R_{\rm s}=4.0 \mpc$ and $\lambda_{\rm th}=0.8$.

The more pronounced trend of sSFR in $R_{\rm s}=4.0 \mpc$ and $\lambda_{\rm th}=0.8$ arises from the interaction of environmental quenching and improved filament identification. $R_{\rm s}=4.0 \mpc$ smooths the density field across more extensive scales, thus eliminating minor fluctuations and highlighting significant filaments. These filaments maintain dynamic stability, while environmental influences, such as ram-pressure stripping and tidal stripping, act effectively. In these well-defined filaments, galaxies located close to spines are subjected to extended periods of quenching processes. 
Briefly, increasing $R_{\rm s}$ or $\lambda_{\rm th}$ better traces physically realistic filaments where environmental impacts are more significant, thus strengthening the observed correlation with the decrease in sSFR around these filaments.

\subsubsection{Galaxy morphology}

\begin{figure}
\includegraphics[width=0.5\textwidth]{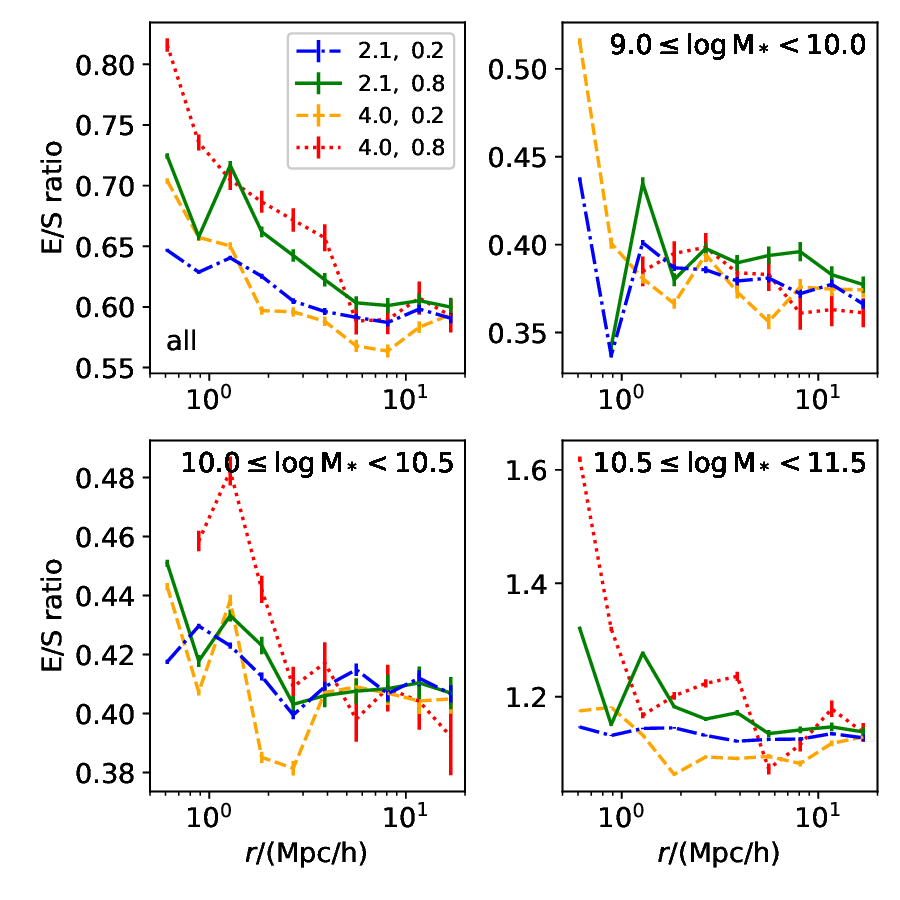}
\caption{Ratios of elliptical to spiral galaxy counts as a function of the distance to the filament spines identified by the COWS method, using varying parameters for $R_{\rm s}=2.1, 4.0\mpc$ and $\lambda_{\rm th}=0.2, 0.8$. The top left panel shows the results for the entire galaxy sample, whereas the remaining panels display findings segmented by galaxy mass categories with $\log (M_*/ (\mstar))$ ranges of $(9,10)$, $(10,10.5)$, and $(10.5,11.5)$, respectively.}
\label{fig:morph}
\end{figure}

The Galaxy Zoo 2 catalogue \citep[GZ2;][]{Willett2013} serves as the basis for categorising galaxies into elliptical and spiral types. Using this catalogue, we have identified $107,230$ elliptical galaxies and $134,024$ spiral galaxies in our NYU-VAGC dataset.

The top left panel of Figure~\ref{fig:morph} presents an illustration showing the ratio of elliptical to spiral galaxy counts (E/S ratio) as a function of their distance from filaments, determined using the COWS method throughout the entire galaxy data set. The remaining panels further detail these findings by galaxy mass with $\log (M_*/ (\mstar))$ defined as intervals of $(9,10)$, $(10,10.5)$, and $(10.5,11.5)$, respectively. For the whole galaxy sample, the E/S ratio tends to increase near the filament spines. 
The trends of the E/S ratio depicted in Figure~\ref{fig:morph} vary based on the parameters $R_s$, $\lambda_{\rm th}$, and the mass of the galaxy.  Galaxies in filaments for $R_s=4.0\mpc$ and $\lambda_{\rm th}=0.8$ show a pronounced increase in the E/S ratio toward their spines, particularly for high-mass galaxies ($\log (M_*/ (\mstar)) \geq 10.5$), consistent with dense filament environments driving morphological quenching. In contrast, trends are weaker or absent for filaments with $R_s=2.1\mpc$ and $\lambda_{\rm th}=0.2$ and low-mass galaxies ($\log (M_*/ (\mstar)) < 10.0$), where environmental effects are less influential. The flat trends observed in these subsets probably indicate that less prominent filaments do not possess sufficient gravitational and thermal energy to strip the gas effectively, resulting in weaker morphological trends. Additionally, low-mass galaxies retain gas reservoirs longer, buffering them against environmental effects. This results in minimal morphological transformation, even near dense filaments.

This result, an increase in the E/S ratio toward filament spines, is qualitatively consistent with the trends reported by \citet{Kuutma2017} and \citet{Okane2024}, although these studies differ in their interpretation of the role of local density. \citet{Kuutma2017} found persistent morphological trends near filaments identified using the Bisous model even after controlling for local density, whereas \citet{Okane2024} reported that the trends vanished for DisPerSE-based filaments when accounting for density. Our results contribute to the growing evidence that filaments influence galaxy morphology while underscoring the need for future studies to jointly model filament geometry and local density.

\subsubsection{Galaxy size in observation}

\begin{figure*}
\includegraphics[width=1.0\textwidth]{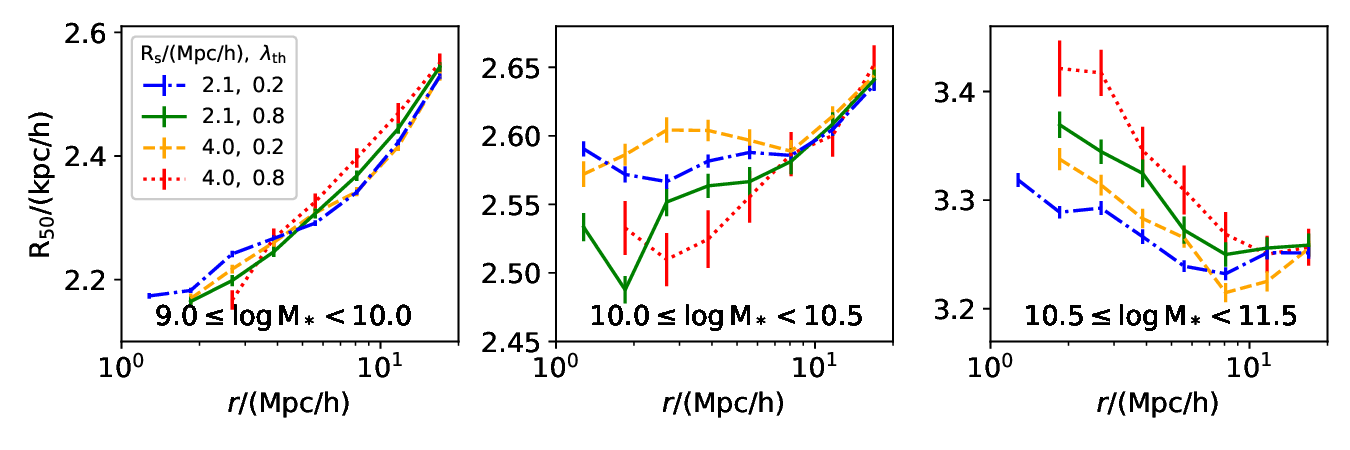}
\caption{Galaxy half-light radii $R_{50}$ as a function of distances from filament spines, identified with the COWS technique for $R_{\rm s}=2.1, 4.0\mpc$ and $\lambda_{\rm th}=0.2, 0.8$, over different mass intervals of $\log (M_*/ (\mstar))$ such as $(9,10)$, $(10,10.5)$, and $(10.5,11.5)$.}
\label{fig:size}
\end{figure*}

The sizes of galaxies are a crucial observational characteristic essential for refining models of how galaxies evolve. Many studies have examined the relationship between galaxy size and stellar mass \citep{Zhang2019}, as well as the scaling relation connecting galaxies to dark-matter haloes \citep{Zhang2022}. 
The established scaling relations between galaxy sizes and stellar masses (e.g., the mass-size relation) show significant scatter in dense environments. By correlating size residuals with filament distance, we can disentangle the roles of internal (e.g., stellar feedback) versus external (e.g., filament-driven accretion) processes in shaping these relations, offering a pathway to refine semi-analytic models. It is intriguing to explore how the size of galaxies is influenced by the large-scale cosmic environment.
Furthermore, it is intriguing to explore how the size of galaxies is influenced by the
large-scale cosmic environment. Consequently, this section investigates the connection between a galaxy's size and its distance to cosmic filaments.

This research employs half-light radii $R_{50}$ of galaxies obtained from the NYU-VAGC catalogue. As reported by \citet{Zhang2022}, the galaxy sizes in the NYU-VAGC catalogue are consistent with those found in the catalogues of \citet{Simard2011} and \citet{Meert2015} for galaxies with similar stellar masses.

Figure~\ref{fig:size} shows how the galaxy size $R_{50}$ correlates with their proximity to the axes of filaments, identified using the COWS technique in various mass ranges: $\log (M_*/ (\mstar))$, including $(9,10)$, $(10,10.5)$, and $(10.5,11.5)$. In the mass intervals of $(9,10)$ or $(10,10.5)$, galaxies generally exhibit smaller sizes near the filament spines. On the other hand, in the $(10.5,11.5)$ mass range, galaxies tend to increase in size as they get closer to the filament axes. A potential reason is that, as low-mass galaxies are pulled into filaments, they tend to become more compact due to tidal stripping. The reduced size of low-mass galaxies near filaments aligns with tidal stripping processes, where galaxies entering dense filamentary environments lose their outer stellar and dark-matter components due to gravitational interactions with neighbouring structures. In contrast, 
high-mass galaxies generally grow in size, primarily as a result of minor mergers \citep{Naab2009} or AGN-driven winds \citep{Coch2023}, which are more common in proximity to filaments. High-mass galaxies near filaments often host active galactic nuclei (AGN). Filament-fed gas inflows trigger AGN activity, generating powerful winds that expel gas from the galactic centre. This feedback suppresses central star formation while redistributing gas to the outer disk, leading to disk expansion.

Analysing galaxies from the LOWZ sample in SDSS DR12, \citet{ChenYC2017} investigated how galaxy size correlates with their distance to cosmic filaments, focussing on galaxies with masses within the range $11.5 \leq \log (M_*/ M_\odot) < 12$. Their findings suggest that smaller galaxies generally have greater mean distances from filaments, consistent with the patterns observed for high-mass galaxies ($\log (M_*/ (\mstar))>10.5$) in this analysis.

\subsection{Halo properties in simulation}\label{sec_halo}

\subsubsection{Subhalo mass}

\begin{figure}
\includegraphics[width=0.5\textwidth]{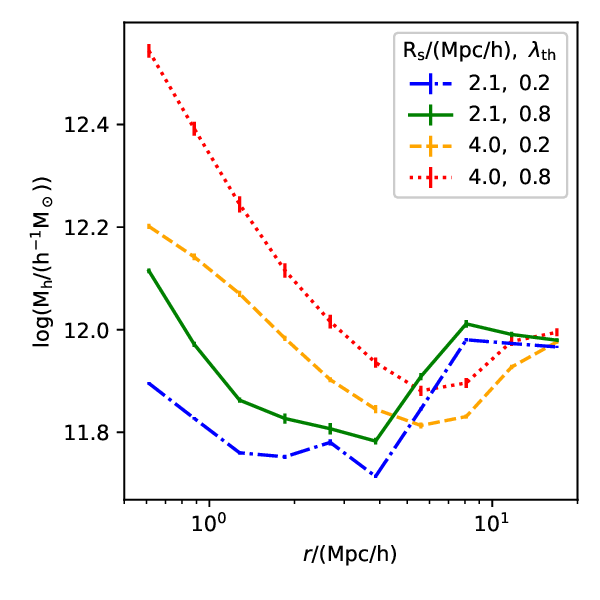}
\caption{Subhalo mass from the ELUCID simulation in relation to their distances from filament spines determined by the COWS technique, using different settings for $R_{\rm s}$ and $\lambda_{\rm th}$. }
\label{fig:subhalo_mass}
\end{figure}

In the ELUCID simulation, the SUBFIND algorithm \citep{Springel2001} is applied to divide a parent FOF halo into several gravitationally bound subhaloes. The largest of these is identified as the central subhalo, and the others are known as satellite subhaloes. To obtain a more precise determination of halo properties such as radius, formation time, spin, and concentration, it is beneficial to use a greater number of particles in larger haloes compared to smaller ones with fewer particles.
Therefore, we mainly focus on central subhaloes with a mass $M_{\rm h}$ greater than $10^{11} \msun$ and containing at least $300$ dark matter particles. 

The mass of the central subhalo, denoted as $M_{\rm h}$, is very close to the virial mass $M_{\rm vir}$ of its host halo. The virial mass is calculated as the total mass within a spherical radius $R_{\rm vir}$, centred on the particle with the lowest potential energy. In the ELUCID simulation, the proportion of $M_{\rm h}$ to $M_{\rm vir}$ for the central subhaloes is determined to be $1.02 \pm 0.08$. In the following analysis, unless otherwise stated, the central subhalo mass $M_{\rm h}$ will be used in this study.

Figure~\ref{fig:subhalo_mass} presents the masses of the central subhaloes in the ELUCID simulation, plotted against their distances from the filament spines identified by the COWS method, with varying parameters for $R_{\rm s}$ and $\lambda_{\rm th}$.
For $R_{\rm s}=4.0 \mpc$, the masses of central subhaloes tend to
decrease as the distance to the nearest filament increases, aligning with the observed pattern that more massive galaxies are situated near filaments. 

For $R_{\rm s}=2.1 \mpc$, the subhalo mass decreases with increasing radial distance $r$, reaching its minimum at approximately twice the smoothing length. Previous studies \citep{Pfeifer2022, YangTianyi2022, Zhang2024} have discovered a valley in the density profiles of filaments. Similarly to these density profiles, the observed valley in halo mass in the vicinity of filaments may stem from the use of the CIC interpolation method within the COWS approach to create the density field with $R_{\rm s} = 2.1 \mpc$. For further details, see Section 3.4 in \citet{Zhang2024}.

\subsubsection{Subhalo formation time}

\begin{figure}
\includegraphics[width=0.5\textwidth]{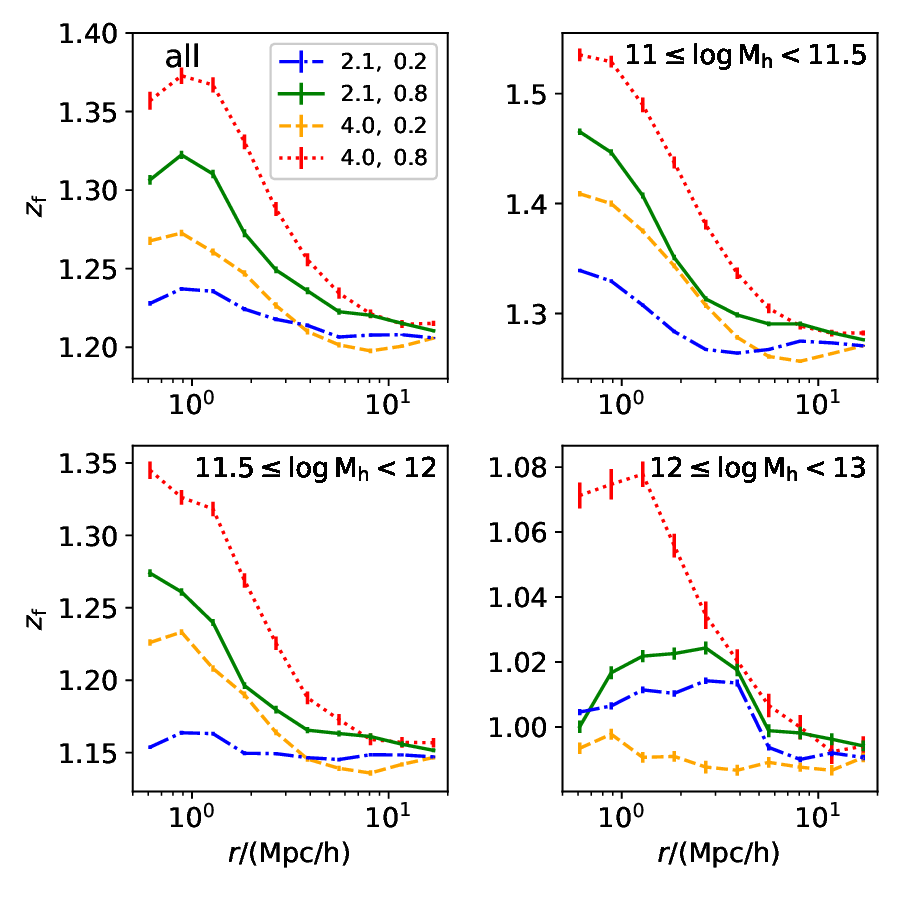}
\caption{Subhalo formation redshift $z_{\rm f}$ as a function of their distance to the filament spines, identified through the COWS method using parameters $R_{\rm s}=2.1, 4.0\mpc$ and $\lambda_{\rm th}=0.2, 0.8$. The top-left panel presents results from our dataset of all central subhaloes with masses greater than $10^{11} \msun$. The other panels show data segregated by subhalo mass ranges for $\log (M_{\rm h}/ (\msun))$, specifically within $(11,11.5)$, $(11.5,12)$, and $(12,13)$.}
\label{fig:age}
\end{figure}

In the ELUCID simulation, merger trees are formed by connecting subhaloes that are detected using the SUBFIND algorithm. The formation redshift of a subhalo, denoted as $z_{\rm f}$, is characterized as the redshift where the main branch progenitor attained half of its maximum mass. To achieve an accurate estimate of the subhalo formation time, our primary focus is on subhaloes comprising a minimum of $300$
dark-matter particles, each having a mass greater than $10^{11} \msun$.

Figure~\ref{fig:age} illustrates the redshift at which central subhalos form, $z_{\rm f}$, as a function of their distances from the cores of filaments, which are identified using the COWS method with parameter values $R_s=2.1, 4.0\mpc$ and $\lambda_{\rm th}=0.2, 0.8$.
The top left panel of Figure~\ref{fig:age} illustrates that for the entire sample of $3,324,498$ central subhaloes with $M_{\rm_h} \geq 10^{11} \msun$, the formation redshifts generally decrease with increasing distance to the filament spines.

To reduce the impact of the subhalo mass on our results, we divide the subhaloes into three distinct mass ranges defined by $\log (M_{\rm h}/ (\msun))$: $(11,11.5)$, $(11.5,12)$, and $(12,13)$. These ranges encompass $2~171~045$, $759~136$, and $354~741$ subhaloes, respectively. As depicted in Figure~\ref{fig:age}, within a specific subhalo mass range, the subhaloes located near the filament spines tend to be older, particularly for $R_s=4.0\mpc$ and $\lambda_{\rm th}=0.8$. The correlation between halo formation redshifts and their proximity to filaments is consistent with patterns noted in galaxies. This is anticipated because galaxies exhibiting redder colours and reduced sSFRs tend to be located closer to filaments.

\subsubsection{Halo size in simulation}
\begin{figure*}
\includegraphics[width=1.0\textwidth]{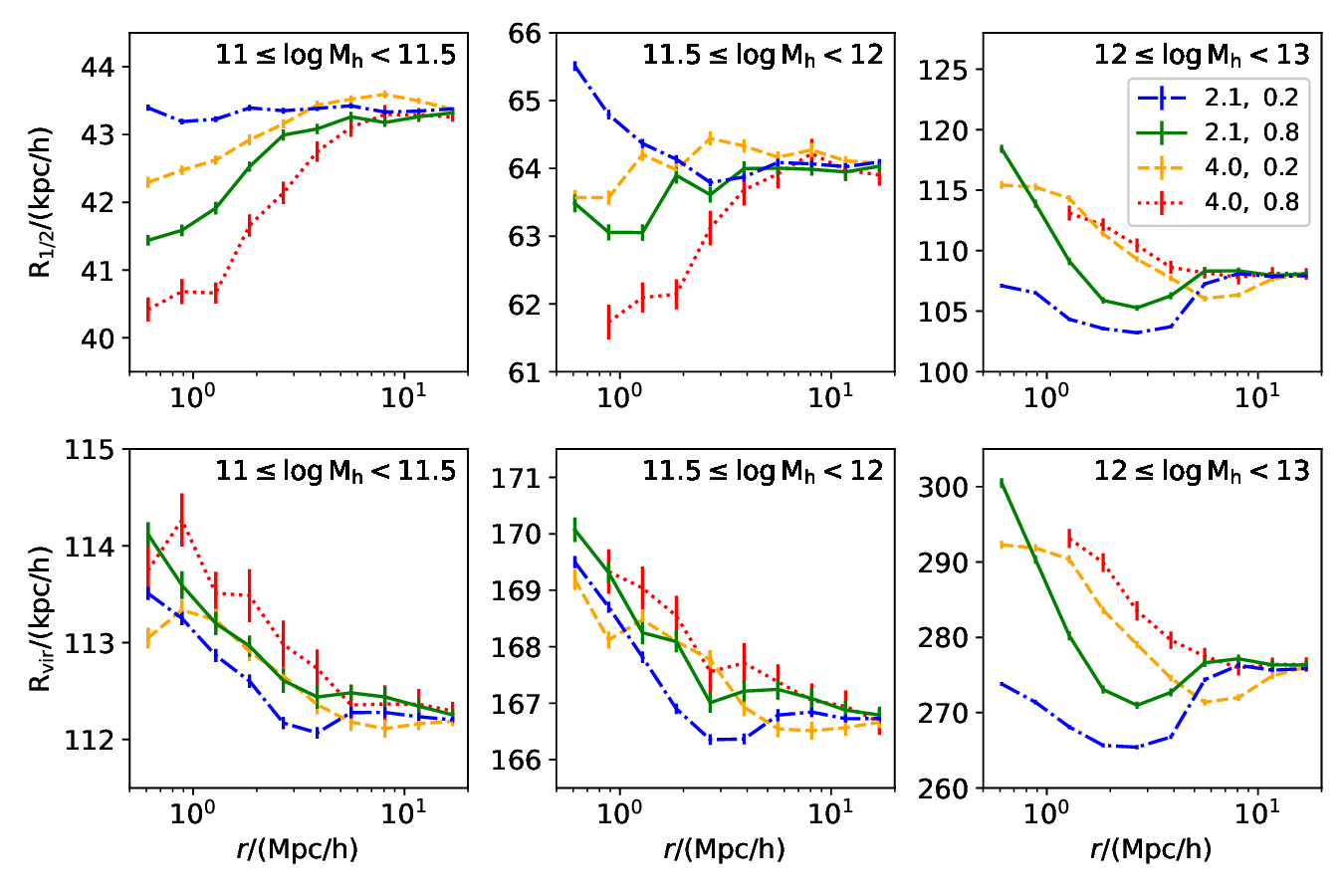}
\caption{Halo radii as a function of the distance to the filament spines that are identified via the COWS method, utilizing different parameters including 
$R_{\rm s}=2.1, 4.0\mpc$ and $\lambda_{\rm th}=0.2, 0.8$, across the mass ranges of $\log (M_{\rm h}/\msun)$ of $(11,11.5)$, $(11.5,12)$, and $(12,13)$. The upper panels show the outcomes for the half-mass radius, whereas the lower panels exhibit the results for the halo virial radius.}
\label{fig:halo_size}
\end{figure*}

This study uses ELUCID simulation to investigate how halo radii, particularly $R_{1/2}$ and $R_{\rm vir}$, vary in relation to their proximity to filament spines. The radius $R_{\rm vir}$ represents the boundary where the mean density of dark-matter haloes is $\Delta_{\rm vir}$ times the universe's critical density \citep{Bryan1998}. The radius $R_{1/2}$ is characterized as the radius enclosing half of the halo's mass $M_{\rm h}$, where $M_{\rm h}=1.02\pm0.08 M_{\rm vir}$ according to the ELUCID simulation framework.

Figure~\ref{fig:halo_size} shows how the halo radii $R_{1/2}$ and $R_{\rm vir}$ vary with the distance to the filament spines identified by the COWS method, considering various parameters such as $R_{\rm s}=2.1, 4.0\mpc$ and $\lambda_{\rm th}=0.2, 0.8$, in the mass ranges of $\log (M_{\rm h}/\msun)$ between $(11,11.5)$, $(11.5,12)$ and $(12,13)$. The lower panels of Figure~\ref{fig:halo_size} illustrate that the virial radii $R_{\rm vir}$ decrease as the distance to the filament spines increases. In contrast, the upper panels show that for low-mass haloes of $11.0 \leq \log (M_{\rm h}/\msun) < 11.5$, the half-mass radii $R_{\rm 1/2}$ increase with increasing distance to the filament spines, particularly for $R_{\rm s} = 4.0\mpc$. 

For low-mass haloes, the decrease in their half-mass radii $R_{\rm 1/2}$ near filaments is similar to the pattern noted in the half-light radii $R_{50}$ of low-mass galaxies with masses in the range $(9.0 \leq \log (M_*/ (\mstar)) < 10.0)$. This is driven by tidal compression and stripping in dense filamentary environments. Filament spines generate strong tidal forces that strip loosely bound outer material, truncating haloes and compacting their mass distribution. This process preferentially removes dark matter from the outskirts of the halo, reducing $R_{\rm 1/2}$ while preserving the core mass.
Generally, tidal stripping affects both dark-matter haloes and their embedded galaxies, suppressing gas reservoirs and star formation in galaxies, leading to compact stellar distributions.

For high-mass haloes, the variation in $R_{1/2}$ in proximity to filaments aligns with the alteration in $R_{\rm vir}$ around filaments.  Massive haloes situated near cosmic filaments primarily expand via mergers and the inflow of gas channelled through the cosmic web, similar to how high-mass galaxies increase in size through mergers. This process of mass accumulation contributes mainly to the outer regions of the halo, resulting in a concurrent expansion of both $R_{\rm vir}$ and $R_{1/2}$.

\subsubsection{Halo concentration and spin}

\begin{figure}
\includegraphics[width=0.5\textwidth]{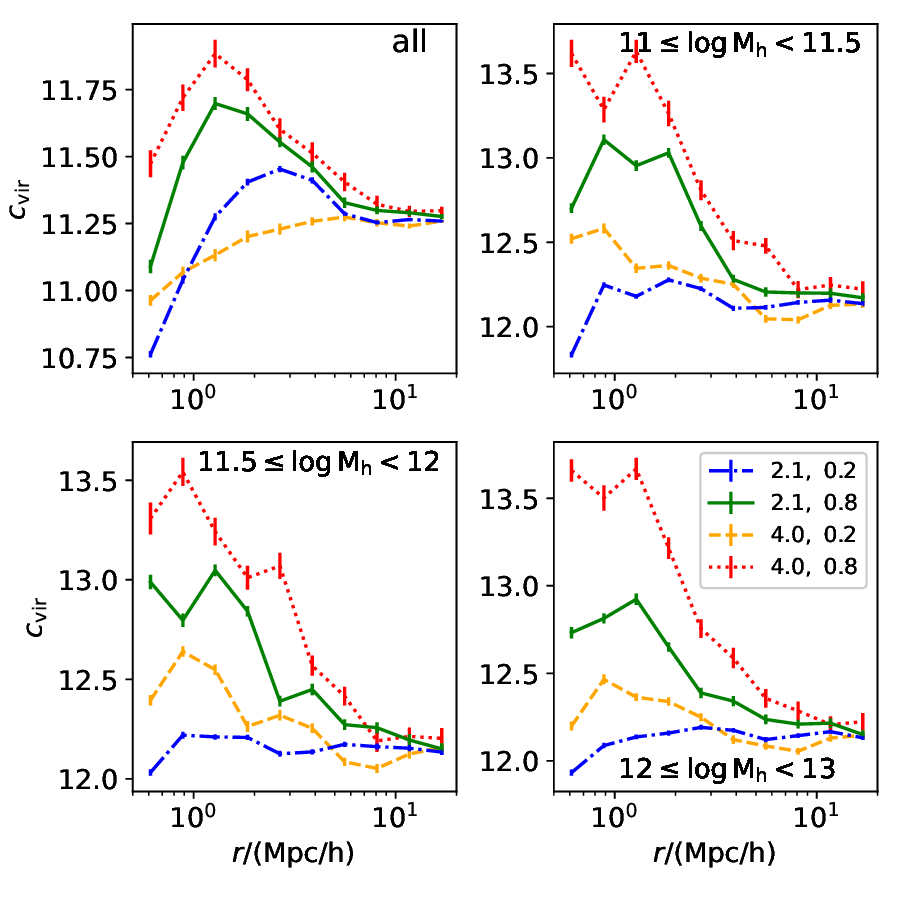}
\caption{Halo concentration $c_{\rm vir}$ as a function of the distance to the filament spines identified by the COWS method, employing different parameters for $R_{\rm s}=2.1, 4.0\mpc$ and $\lambda_{\rm th}=0.2, 0.8$. The top-left panel illustrates data from our sample of all central subhaloes with masses exceeding $10^{11} \msun$. The remaining panels display data categorized by subhalo mass brackets for $\log (M_{\rm h}/ (\msun))$, specifically within the intervals $(11,11.5)$, $(11.5,12)$, and $(12,13)$.}
\label{fig:con}
\end{figure}

\begin{figure}
\includegraphics[width=0.5\textwidth]{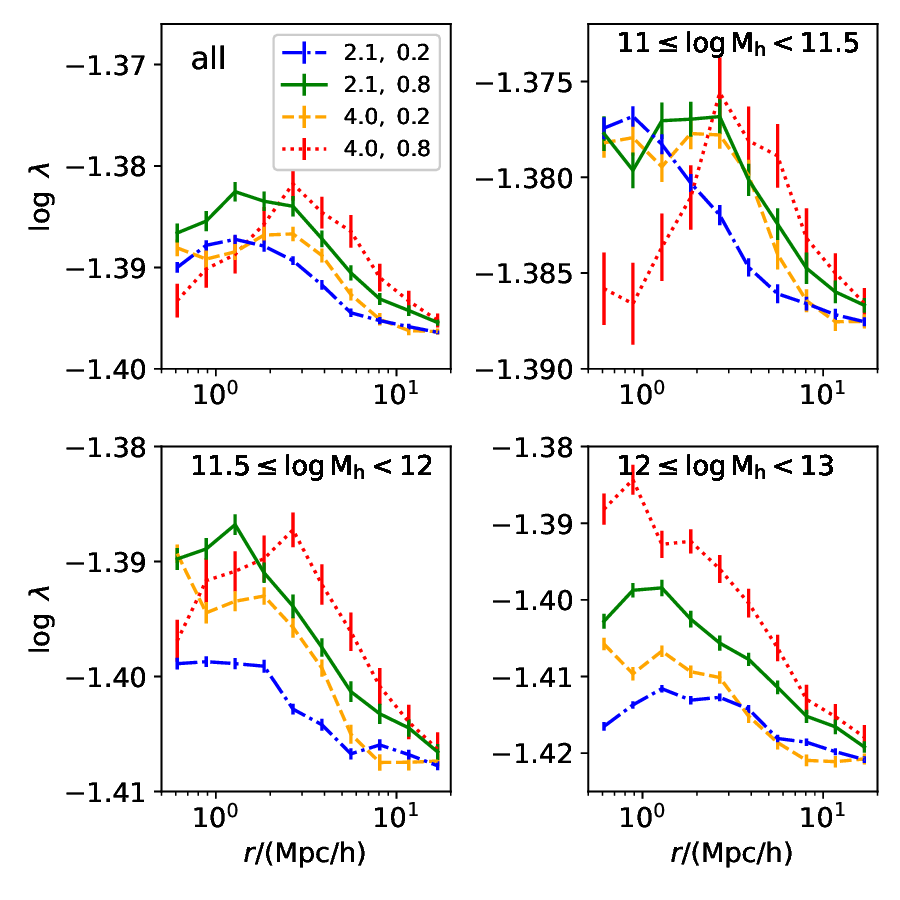}
\caption{Similar to Figure~\ref{fig:con}, but for the halo spin parameter $\lambda$.}
\label{fig:spin}
\end{figure}

This section investigates how halo concentration and spin vary in relation to their distance to filament spines, based on data obtained from the ELUCID simulation.

The concentration values $c_{\rm vir}$ are derived from dark matter haloes based on the ELUCID simulation, as described in Equation~\ref{eqn:cvir}. Figure~\ref{fig:con} presents the variation of the halo concentration $c_{\rm vir}$ relative to the proximity to the filament cores, determined using the COWS method with different settings for $R_{\rm s}=2.1, 4.0\mpc$ and $\lambda_{\rm th}=0.2, 0.8$. 
In general, as the halos approach the filaments, their concentration increases, particularly when $R_{\rm s}=4.0\mpc$ and $\lambda_{\rm th}=0.8$. However, this pattern is not observed for $R_{\rm s}=2.1\mpc$ and $\lambda_{\rm th}=0.2$.

Using two distinct collections of cosmological hydrodynamical simulations, \citet{Jiang2019} demonstrated that the ratio of the size of a galaxy to the virial radius of the halo is expressed as $c_{\rm vir}^{-0.7}$. This result suggests an inverse correlation between galaxy size and halo concentration. In the case of galaxies that reside in low-mass haloes, an increase in halo concentration close to filaments could lead to reduced sizes of such low-mass galaxies near these filaments.
For galaxies in high-mass haloes, galaxy mergers may be the main driver of the size evolution \citep{Khochfar2006, Cooper2012, Figueira2024}.

In this study, the spin values $\lambda$ are derived from Equation~\ref{eqn:spin} with the use of subhaloes within the ELUCID simulation. Figure~\ref{fig:con} depicts the halo spin parameter $\lambda$ as a function of the distance to the filament spines determined using the COWS method, using varying settings for $R_{\rm s}=2.1, 4.0\mpc$ and $\lambda_{\rm th}=0.2, 0.8$. The panel in the upper left presents data from our collection of all central subhaloes with masses greater than $10^{11} \msun$. The other panels illustrate data segmented by subhalo mass ranges for $\log (M_{\rm h}/ (\msun))$, covering the intervals $(11,11.5)$, $(11.5,12)$, and $(12,13)$.

In general, as the distance to the filament axes increases, the spin values tend to decrease. This pattern is consistent with the findings of \citet{XueWX2024}, who analyzed the rotational speeds of haloes by studying \ion{H}{I} gas-rich galaxies observed in the Arecibo Legacy Fast Alfa Survey. Their research revealed that haloes with higher spin values tend to be situated nearer to filaments.

The tendency of high-spin haloes to reside near filaments arises from the interplay between filament-driven angular momentum transfer and anisotropic mass accretion. 

During the stages of structure formation, haloes acquire angular momentum through tidal torquing, where misalignments between the halo inertia tensor and the local tidal field generate spin. Filaments are regions of high tidal shear due to their elongated geometry and dense matter distribution. This amplifies the tidal torque, increasing the acquisition of angular momentum compared to isotropic environments. 

Filaments channel gas and dark matter flows preferentially along their axes, driving coherent accretion onto haloes. Gas streaming along filaments retains the angular momentum of large-scale flows, which is transferred to haloes during accretion. 
While major mergers randomise the spin of the halo, minor mergers, which dominate in filament environments, can amplify the spin. Minor mergers along filaments deliver satellites with aligned trajectories, transferring orbital angular momentum to the spin of the halo.

Generally, filaments act as cosmic highways for angular momentum, channelling coherent flows that spin up haloes through tidal torquing, anisotropic accretion, and minor mergers. This explains the observed spin-distance trend and aligns with the $\Lambda$-CDM predictions of the web-regulated galaxy-halo evolution.

\section{Summary}\label{sec_summary}

In this study, we examine the properties of galaxies and haloes in terms of their distance to cosmic filaments, using galaxies from SDSS DR7 and haloes in the constrained ELUCID simulation. The characteristics of galaxies that are being analyzed encompass stellar mass, colour, specific star formation rate (sSFR), size, and the elliptical-to-spiral ratio (E/S ratio). In terms of haloes, our study focuses on attributes such as mass, age, radius, spin, and concentration.

The structure of the cosmic web, which consists of knots, filaments, sheets, and voids, is classified through the Hessian matrix related to the density field, as detailed in Equation~\ref{eqn:hessian}. 
In order to examine the properties of galaxies or haloes surrounding filaments, the filament axes are determined using the medial axis thinning technique implemented in the COWS method \citep{Pfeifer2022}. As illustrated in Figure~\ref{fig:slice}, the medial axes effectively outline the spines of cosmic filaments, which are prominently shaped by the arrangement of dark matter particles.
Our validation process confirmed that the filament axes obtained using the medial axis thinning technique are well aligned with the eigenvector of the Hessian matrix corresponding to the orientation of the filament, thus supporting the filament axes identification method \citep{Zhang2024}.

Although the medial-axis thinning algorithm operates without any adjustable parameters, the method to classify the cosmic web depends on two specific variables: the smoothing length $R_{\rm s}$ and the threshold $\lambda_{\rm th}$. We verified our results with several filament catalogues using different values of $R_{\rm s}$ or $\lambda_{\rm th}$, and in most cases, the main conclusions are consistent in various parameters.

However, filaments with reduced lengths and thinner shapes, specified by parameters $R_{\rm s} = 2.1 \mpc, \lambda_{\rm th} = 0.2$, show a weak trend of increase or decrease, as represented by the blue dot-dashed lines. The weak increase or decrease in the tread for $R_{\rm s} = 2.1 \mpc, \lambda_{\rm th} = 0.2$ resembles the density reconstruction impact as described in \citep{Hasan2024}, where the use of DTFE density reconstruction to study galaxies indicates a decrease in star formation in galaxies near filaments. In contrast, filaments identified using the MCPM method appear to exert minimal impact on star formation activities. 

As noted in \citet{Hasan2024}, the MCPM density reconstruction offers a significantly detailed depiction of the cosmic density field, allowing for the detection of less prominent filaments that the DTFE method misses. This scenario resembles our approach, where a reduction in $R_{\rm s}$ improves the cosmological density resolution, allowing the identification of subtler filaments and producing a flatter trend for smaller $R_{\rm s}$ values.

This research sets itself apart with distinctive methodological techniques and constrained simulations that enhance our understanding of the influence of cosmic filaments on galaxy evolution. Unlike many studies that use algorithms like DisPerSE or NEXUS (which often focus on density ridges or topological persistence), this work employs the COWS method to define filaments as medial axes (skeletons) of the cosmic web. The study uniquely pairs SDSS observations with the ELUCID simulation, a constrained dark-matter-only simulation reconstructed to match the observed local universe (for example, galaxy groups and filaments). Unlike random cosmological simulations, the alignment of ELUCID with the real universe minimises sampling bias, enabling direct statistically robust comparisons between simulations and observations. The summary of our findings is presented below.

\begin{enumerate}
\item [(i)]
As galaxies or central subhaloes approach closer to the filaments, their masses tend to increase, especially when higher values for the parameters $R_{\rm s}$ or $\lambda_{\rm th}$ are used in the filament classification technique.

\item [(ii)]
Galaxies that are situated close to the filament axes generally display a redder colour and usually present lower specific star formation rates. Subhaloes within a certain mass range are typically older when located near the core areas of filaments, particularly for $R_{\rm s} = 4.0 \mpc$ or $\lambda_{\rm th}=0.8$.

\item [(iii)]
The study uses the Galaxy Zoo 2 catalogue to classify galaxies as elliptical or spiral. The findings indicate a higher proportion of elliptical galaxies compared to spirals closer to the filament axes, particularly evident for $R_{\rm s} = 4.0 \mpc$ or $\lambda_{\rm th}=0.8$.

\item [(iv)]
Galaxies with lower mass are typically smaller in size when located near the filament spines. In contrast, galaxies of higher mass, specifically those with $\log (M_*/ (\mstar)) \geq 10.5$, tend to become larger as they approach the filament axes. The pattern found in the half-mass radii $R_{1/2}$ of haloes positioned close to filaments is similar to the trend observed in the half-light radii $R_{50}$ of galaxies.

\item [(v)]
In general, as halos approach filaments, their concentration or spin tends to increase. This suggests that the cosmic web exerts a notable influence on the dynamics of halos.

\end{enumerate}

In conclusion, this research uses the COWS method to detect cosmic web structures, revealing that the properties of galaxies or halos are significantly affected by their distance to cosmic filaments. The results support the notion that the large-scale structure of the universe, characterised by cosmic filaments, plays a vital role in shaping galaxy and halo properties, which are consistent with patterns noted in previous research. Motivated by this, upcoming studies will investigate the ways in which galaxy or halo characteristics are affected and evolved along the evolution of the cosmic web, from cosmic void, sheet, to filament, and then clusters.

It is important to note that different filament detection algorithms may produce a variety of distinct results. For example, the COWS method evaluates the properties of galaxies beginning at a distance of $0.5 \mpc$ from the filament's axis. This approach uses a grid structure to establish the axis of the filament, with the finest resolution being equivalent to the grid's size. However, using the axis determined by DisPerSE, calculations can be performed efficiently within $0.5 \mpc$. In particular, the DisPerSE method indicates that the boundary of the filament, as described in \citep{WangWei2024}, is roughly $1 {\rm Mpc}$. Besides the difficulty of defining the filament axis, there are other factors that could be significant. In the COWS approach, altering the parameters $R_{\rm s}$ and $\lambda_{\rm th}$ in the filament catalogue results in variations in the distribution of characteristics among the galaxies surrounding the filaments. Alternative methods employed to detect filaments are also likely to affect how galaxy features are distributed around these filaments. In forthcoming studies, we plan to explore how various filament detection techniques influence galaxy properties.

\section*{Acknowledgements}

This research is funded by various grants, including the National SKA Program of China (grant No. 2020SKA0110100), the National Natural Science Foundation of China (Nos. 12273088), and the National Key R\&D Programme of China (2023YFA1607800, 2023YFA1607804). Additional support comes from the CSST project (No. CMS-CSST-2021-A02), the CAS Project for Young Scientists in Basic Research (No. YSBR-092), Fundamental Research Funds for Central Universities, the 111 project (No. B20019) and the Shanghai Natural Science Foundation (grant No.19ZR1466800, 23JC1410200, ZJ20223-ZD-003). PW acknowledge financial support by the NSFC (No. 12473009), and also sponsored by Shanghai Rising-Star Program (No.24QA2711100).

This work is also supported by the High-Performance Computing Resource in the Core Facility for Advanced Research Computing at Shanghai Astronomical Observatory.

This work used {\tt Astropy:} a community-developed core Python package
and an ecosystem of tools and resources for astronomy \citep{Astropy2013, 
Astropy2018, Astropy2022}. 

Funding for the Sloan Digital Sky Survey IV has been provided by the
Alfred P. Sloan Foundation, the U.S. Department of Energy Office of
Science, and the Participating Institutions. SDSS acknowledges support
and resources from the Centre for High-Performance Computing at the
University of Utah. The SDSS website is www.sdss.org.

SDSS is managed by the Astrophysical Research Consortium for the
Participating Institutions of the SDSS Collaboration including the
Brazilian Participation Group, the Carnegie Institution for Science,
Carnegie Mellon University, the Chilean Participation Group, the
French Participation Group, Harvard-Smithsonian Center for
Astrophysics, Instituto de Astrof{\'i}sica de Canarias, The Johns
Hopkins University, Kavli Institute for the Physics and Mathematics of
the Universe (IPMU)/University of Tokyo, Lawrence Berkeley National
Laboratory, Leibniz Institut f{\"u}r Astrophysik Potsdam (AIP),
Max-Planck-Institut f{\"u}r Astronomie (MPIA Heidelberg),
Max-Planck-Institut f{\"u}r Astrophysik (MPA Garching),
Max-Planck-Institut f{\"u}r Extraterrestrische Physik (MPE), National
Astronomical Observatories of China, New Mexico State University, New
York University, University of Notre Dame, Observat{\'o}rio Nacional/
MCTI, The Ohio State University, Pennsylvania State University,
Shanghai Astronomical Observatory, United Kingdom Participation Group,
Universidad Nacional Aut{\'o}noma de M{\'e}xico, University of
Arizona, University of Colorado Boulder, University of Oxford,
University of Portsmouth, University of Utah, University of Virginia,
University of Washington, University of Wisconsin, Vanderbilt
University, and Yale University.

\section*{Data availability}
The data underlying this article will be shared on reasonable request with
the corresponding author.

\bibliographystyle{mnras}
\bibliography{bibliography}

\begin{thebibliography}{}
\makeatletter
\relax
\def\mn@urlcharsother{\let\do\@makeother \do\$\do\&\do\#\do\^\do\_\do\%\do\~}
\def\mn@doi{\begingroup\mn@urlcharsother \@ifnextchar [ {\mn@doi@} {\mn@doi@[]}}
\def\mn@doi@[#1]#2{\def\@tempa{#1}\ifx\@tempa\@empty \href {http://dx.doi.org/#2} {doi:#2}\else \href {http://dx.doi.org/#2} {#1}\fi \endgroup}
\def\mn@eprint#1#2{\mn@eprint@#1:#2::\@nil}
\def\mn@eprint@arXiv#1{\href {http://arxiv.org/abs/#1} {{\tt arXiv:#1}}}
\def\mn@eprint@dblp#1{\href {http://dblp.uni-trier.de/rec/bibtex/#1.xml} {dblp:#1}}
\def\mn@eprint@#1:#2:#3:#4\@nil{\def\@tempa {#1}\def\@tempb {#2}\def\@tempc {#3}\ifx \@tempc \@empty \let \@tempc \@tempb \let \@tempb \@tempa \fi \ifx \@tempb \@empty \def\@tempb {arXiv}\fi \@ifundefined {mn@eprint@\@tempb}{\@tempb:\@tempc}{\expandafter \expandafter \csname mn@eprint@\@tempb\endcsname \expandafter{\@tempc}}}

\bibitem[\protect\citeauthoryear{{Abazajian} et~al.,}{{Abazajian} et~al.}{2009}]{Abazajian2009}
{Abazajian} K.~N.,  et~al., 2009, \mn@doi [\apjs] {10.1088/0067-0049/182/2/543}, \href {https://ui.adsabs.harvard.edu/abs/2009ApJS..182..543A} {182, 543}

\bibitem[\protect\citeauthoryear{{Alpaslan} et~al.,}{{Alpaslan} et~al.}{2014}]{Alpaslan2014}
{Alpaslan} M.,  et~al., 2014, \mn@doi [\mnras] {10.1093/mnras/stt2136}, \href {https://ui.adsabs.harvard.edu/abs/2014MNRAS.438..177A} {438, 177}

\bibitem[\protect\citeauthoryear{{Alpaslan} et~al.,}{{Alpaslan} et~al.}{2015}]{Alpaslan2015}
{Alpaslan} M.,  et~al., 2015, \mn@doi [\mnras] {10.1093/mnras/stv1176}, \href {https://ui.adsabs.harvard.edu/abs/2015MNRAS.451.3249A} {451, 3249}

\bibitem[\protect\citeauthoryear{{Astropy Collaboration} et~al.,}{{Astropy Collaboration} et~al.}{2013}]{Astropy2013}
{Astropy Collaboration} et~al., 2013, \mn@doi [\aap] {10.1051/0004-6361/201322068}, \href {https://ui.adsabs.harvard.edu/abs/2013A&A...558A..33A} {558, A33}

\bibitem[\protect\citeauthoryear{{Astropy Collaboration} et~al.,}{{Astropy Collaboration} et~al.}{2018}]{Astropy2018}
{Astropy Collaboration} et~al., 2018, \mn@doi [\aj] {10.3847/1538-3881/aabc4f}, \href {https://ui.adsabs.harvard.edu/abs/2018AJ....156..123A} {156, 123}

\bibitem[\protect\citeauthoryear{{Astropy Collaboration} et~al.,}{{Astropy Collaboration} et~al.}{2022}]{Astropy2022}
{Astropy Collaboration} et~al., 2022, \mn@doi [\apj] {10.3847/1538-4357/ac7c74}, \href {https://ui.adsabs.harvard.edu/abs/2022ApJ...935..167A} {935, 167}

\bibitem[\protect\citeauthoryear{{Bell}, {McIntosh}, {Katz}  \& {Weinberg}}{{Bell} et~al.}{2003}]{Bell2003}
{Bell} E.~F.,  {McIntosh} D.~H.,  {Katz} N.,   {Weinberg} M.~D.,  2003, \mn@doi [\apjs] {10.1086/378847}, \href {https://ui.adsabs.harvard.edu/abs/2003ApJS..149..289B} {149, 289}

\bibitem[\protect\citeauthoryear{{Blanton} et~al.,}{{Blanton} et~al.}{2001}]{Blanton2001}
{Blanton} M.~R.,  et~al., 2001, \mn@doi [\aj] {10.1086/320405}, \href {https://ui.adsabs.harvard.edu/abs/2001AJ....121.2358B} {121, 2358}

\bibitem[\protect\citeauthoryear{{Blanton} et~al.,}{{Blanton} et~al.}{2005}]{Blanton2005}
{Blanton} M.~R.,  et~al., 2005, \mn@doi [\aj] {10.1086/429803}, \href {https://ui.adsabs.harvard.edu/abs/2005AJ....129.2562B} {129, 2562}

\bibitem[\protect\citeauthoryear{{Bond}, {Kofman}  \& {Pogosyan}}{{Bond} et~al.}{1996}]{Bond1996}
{Bond} J.~R.,  {Kofman} L.,   {Pogosyan} D.,  1996, \mn@doi [\nat] {10.1038/380603a0}, \href {https://ui.adsabs.harvard.edu/abs/1996Natur.380..603B} {380, 603}

\bibitem[\protect\citeauthoryear{{Bonjean}, {Aghanim}, {Douspis}, {Malavasi}  \& {Tanimura}}{{Bonjean} et~al.}{2020}]{Bonjean2020}
{Bonjean} V.,  {Aghanim} N.,  {Douspis} M.,  {Malavasi} N.,   {Tanimura} H.,  2020, \mn@doi [\aap] {10.1051/0004-6361/201937313}, \href {https://ui.adsabs.harvard.edu/abs/2020A&A...638A..75B} {638, A75}

\bibitem[\protect\citeauthoryear{{Bryan} \& {Norman}}{{Bryan} \& {Norman}}{1998}]{Bryan1998}
{Bryan} G.~L.,  {Norman} M.~L.,  1998, \mn@doi [\apj] {10.1086/305262}, \href {https://ui.adsabs.harvard.edu/abs/1998ApJ...495...80B} {495, 80}

\bibitem[\protect\citeauthoryear{{Bulichi}, {Dav{\'e}}  \& {Kraljic}}{{Bulichi} et~al.}{2024}]{Bulichi2024}
{Bulichi} T.-E.,  {Dav{\'e}} R.,   {Kraljic} K.,  2024, \mn@doi [\mnras] {10.1093/mnras/stae667}, \href {https://ui.adsabs.harvard.edu/abs/2024MNRAS.529.2595B} {529, 2595}

\bibitem[\protect\citeauthoryear{{Carr{\'o}n Duque}, {Migliaccio}, {Marinucci}  \& {Vittorio}}{{Carr{\'o}n Duque} et~al.}{2022}]{Carron2022}
{Carr{\'o}n Duque} J.,  {Migliaccio} M.,  {Marinucci} D.,   {Vittorio} N.,  2022, \mn@doi [\aap] {10.1051/0004-6361/202141538}, \href {https://ui.adsabs.harvard.edu/abs/2022A&A...659A.166C} {659, A166}

\bibitem[\protect\citeauthoryear{{Cautun}, {van de Weygaert}  \& {Jones}}{{Cautun} et~al.}{2013}]{Cautun2013}
{Cautun} M.,  {van de Weygaert} R.,   {Jones} B. J.~T.,  2013, \mn@doi [\mnras] {10.1093/mnras/sts416}, \href {https://ui.adsabs.harvard.edu/abs/2013MNRAS.429.1286C} {429, 1286}

\bibitem[\protect\citeauthoryear{{Cautun}, {van de Weygaert}, {Jones}  \& {Frenk}}{{Cautun} et~al.}{2014}]{Cautun2014}
{Cautun} M.,  {van de Weygaert} R.,  {Jones} B. J.~T.,   {Frenk} C.~S.,  2014, \mn@doi [\mnras] {10.1093/mnras/stu768}, \href {https://ui.adsabs.harvard.edu/abs/2014MNRAS.441.2923C} {441, 2923}

\bibitem[\protect\citeauthoryear{{Chang}, {van der Wel}, {da Cunha}  \& {Rix}}{{Chang} et~al.}{2015}]{Chang2015}
{Chang} Y.-Y.,  {van der Wel} A.,  {da Cunha} E.,   {Rix} H.-W.,  2015, \mn@doi [\apjs] {10.1088/0067-0049/219/1/8}, \href {https://ui.adsabs.harvard.edu/abs/2015ApJS..219....8C} {219, 8}

\bibitem[\protect\citeauthoryear{{Chen} et~al.,}{{Chen} et~al.}{2017}]{ChenYC2017}
{Chen} Y.-C.,  et~al., 2017, \mn@doi [\mnras] {10.1093/mnras/stw3127}, \href {https://ui.adsabs.harvard.edu/abs/2017MNRAS.466.1880C} {466, 1880}

\bibitem[\protect\citeauthoryear{{Chen}, {Mo}, {Li}, {Wang}, {Yang}, {Zhou}  \& {Zhang}}{{Chen} et~al.}{2019}]{Chen2019}
{Chen} Y.,  {Mo} H.~J.,  {Li} C.,  {Wang} H.,  {Yang} X.,  {Zhou} S.,   {Zhang} Y.,  2019, \mn@doi [\apj] {10.3847/1538-4357/ab0208}, \href {https://ui.adsabs.harvard.edu/abs/2019ApJ...872..180C} {872, 180}

\bibitem[\protect\citeauthoryear{{Cochrane} et~al.,}{{Cochrane} et~al.}{2023}]{Coch2023}
{Cochrane} R.~K.,  et~al., 2023, \mn@doi [\mnras] {10.1093/mnras/stad1528}, \href {https://ui.adsabs.harvard.edu/abs/2023MNRAS.523.2409C} {523, 2409}

\bibitem[\protect\citeauthoryear{{Conrado} et~al.,}{{Conrado} et~al.}{2024}]{Conrado2024}
{Conrado} A.~M.,  et~al., 2024, \mn@doi [\aap] {10.1051/0004-6361/202449414}, \href {https://ui.adsabs.harvard.edu/abs/2024A&A...687A..98C} {687, A98}

\bibitem[\protect\citeauthoryear{{Cooper} et~al.,}{{Cooper} et~al.}{2012}]{Cooper2012}
{Cooper} M.~C.,  et~al., 2012, \mn@doi [\mnras] {10.1111/j.1365-2966.2011.19938.x}, \href {https://ui.adsabs.harvard.edu/abs/2012MNRAS.419.3018C} {419, 3018}

\bibitem[\protect\citeauthoryear{{Curtis}, {McDonough}  \& {Brainerd}}{{Curtis} et~al.}{2024}]{Curtis2024}
{Curtis} O.,  {McDonough} B.,   {Brainerd} T.~G.,  2024, \mn@doi [\apj] {10.3847/1538-4357/ad18b4}, \href {https://ui.adsabs.harvard.edu/abs/2024ApJ...962...58C} {962, 58}

\bibitem[\protect\citeauthoryear{{DESI Collaboration} et~al.,}{{DESI Collaboration} et~al.}{2024}]{DESI2024}
{DESI Collaboration} et~al., 2024, \mn@doi [\aj] {10.3847/1538-3881/ad3217}, \href {https://ui.adsabs.harvard.edu/abs/2024AJ....168...58D} {168, 58}

\bibitem[\protect\citeauthoryear{{Darvish}, {Sobral}, {Mobasher}, {Scoville}, {Best}, {Sales}  \& {Smail}}{{Darvish} et~al.}{2014}]{Darvish2014}
{Darvish} B.,  {Sobral} D.,  {Mobasher} B.,  {Scoville} N.~Z.,  {Best} P.,  {Sales} L.~V.,   {Smail} I.,  2014, \mn@doi [\apj] {10.1088/0004-637X/796/1/51}, \href {https://ui.adsabs.harvard.edu/abs/2014ApJ...796...51D} {796, 51}

\bibitem[\protect\citeauthoryear{{Davis}, {Efstathiou}, {Frenk}  \& {White}}{{Davis} et~al.}{1985}]{Davis1985}
{Davis} M.,  {Efstathiou} G.,  {Frenk} C.~S.,   {White} S.~D.~M.,  1985, \mn@doi [\apj] {10.1086/163168}, \href {https://ui.adsabs.harvard.edu/abs/1985ApJ...292..371D} {292, 371}

\bibitem[\protect\citeauthoryear{{Dom{\'\i}nguez-G{\'o}mez} et~al.,}{{Dom{\'\i}nguez-G{\'o}mez} et~al.}{2023}]{Dom2023}
{Dom{\'\i}nguez-G{\'o}mez} J.,  et~al., 2023, \mn@doi [\aap] {10.1051/0004-6361/202346884}, \href {https://ui.adsabs.harvard.edu/abs/2023A&A...680A.111D} {680, A111}

\bibitem[\protect\citeauthoryear{{Donnan}, {Tojeiro}  \& {Kraljic}}{{Donnan} et~al.}{2022}]{Donnan2022}
{Donnan} C.~T.,  {Tojeiro} R.,   {Kraljic} K.,  2022, \mn@doi [Nature Astronomy] {10.1038/s41550-022-01619-w}, \href {https://ui.adsabs.harvard.edu/abs/2022NatAs...6..599D} {6, 599}

\bibitem[\protect\citeauthoryear{{Figueira}, {Siudek}, {Pollo}, {Krywult}, {Vergani}, {Bolzonella}, {Cucciati}  \& {Iovino}}{{Figueira} et~al.}{2024}]{Figueira2024}
{Figueira} M.,  {Siudek} M.,  {Pollo} A.,  {Krywult} J.,  {Vergani} D.,  {Bolzonella} M.,  {Cucciati} O.,   {Iovino} A.,  2024, \mn@doi [\aap] {10.1051/0004-6361/202347774}, \href {https://ui.adsabs.harvard.edu/abs/2024A&A...687A.117F} {687, A117}

\bibitem[\protect\citeauthoryear{{Hahn}, {Porciani}, {Carollo}  \& {Dekel}}{{Hahn} et~al.}{2007}]{Hahn2007}
{Hahn} O.,  {Porciani} C.,  {Carollo} C.~M.,   {Dekel} A.,  2007, \mn@doi [\mnras] {10.1111/j.1365-2966.2006.11318.x}, \href {https://ui.adsabs.harvard.edu/abs/2007MNRAS.375..489H} {375, 489}

\bibitem[\protect\citeauthoryear{{Hasan} et~al.,}{{Hasan} et~al.}{2023}]{Hasan2023}
{Hasan} F.,  et~al., 2023, \mn@doi [\apj] {10.3847/1538-4357/acd11c}, \href {https://ui.adsabs.harvard.edu/abs/2023ApJ...950..114H} {950, 114}

\bibitem[\protect\citeauthoryear{{Hasan} et~al.,}{{Hasan} et~al.}{2024}]{Hasan2024}
{Hasan} F.,  et~al., 2024, \mn@doi [\apj] {10.3847/1538-4357/ad4ee2}, \href {https://ui.adsabs.harvard.edu/abs/2024ApJ...970..177H} {970, 177}

\bibitem[\protect\citeauthoryear{{Hoosain} et~al.,}{{Hoosain} et~al.}{2024}]{Hoosain2024}
{Hoosain} M.,  et~al., 2024, \mn@doi [\mnras] {10.1093/mnras/stae174}, \href {https://ui.adsabs.harvard.edu/abs/2024MNRAS.528.4139H} {528, 4139}

\bibitem[\protect\citeauthoryear{{Hoyle}, {Rojas}, {Vogeley}  \& {Brinkmann}}{{Hoyle} et~al.}{2005}]{Hoyle2005}
{Hoyle} F.,  {Rojas} R.~R.,  {Vogeley} M.~S.,   {Brinkmann} J.,  2005, \mn@doi [\apj] {10.1086/427176}, \href {https://ui.adsabs.harvard.edu/abs/2005ApJ...620..618H} {620, 618}

\bibitem[\protect\citeauthoryear{{Hoyle}, {Vogeley}  \& {Pan}}{{Hoyle} et~al.}{2012}]{Hoyle2012}
{Hoyle} F.,  {Vogeley} M.~S.,   {Pan} D.,  2012, \mn@doi [\mnras] {10.1111/j.1365-2966.2012.21943.x}, \href {https://ui.adsabs.harvard.edu/abs/2012MNRAS.426.3041H} {426, 3041}

\bibitem[\protect\citeauthoryear{{Jiang} et~al.,}{{Jiang} et~al.}{2019}]{Jiang2019}
{Jiang} F.,  et~al., 2019, \mn@doi [\mnras] {10.1093/mnras/stz1952}, \href {https://ui.adsabs.harvard.edu/abs/2019MNRAS.488.4801J} {488, 4801}

\bibitem[\protect\citeauthoryear{{Khochfar} \& {Silk}}{{Khochfar} \& {Silk}}{2006}]{Khochfar2006}
{Khochfar} S.,  {Silk} J.,  2006, \mn@doi [\apjl] {10.1086/507768}, \href {https://ui.adsabs.harvard.edu/abs/2006ApJ...648L..21K} {648, L21}

\bibitem[\protect\citeauthoryear{{Kleiner}, {Pimbblet}, {Jones}, {Koribalski}  \& {Serra}}{{Kleiner} et~al.}{2017}]{Kleiner2017}
{Kleiner} D.,  {Pimbblet} K.~A.,  {Jones} D.~H.,  {Koribalski} B.~S.,   {Serra} P.,  2017, \mn@doi [\mnras] {10.1093/mnras/stw3328}, \href {https://ui.adsabs.harvard.edu/abs/2017MNRAS.466.4692K} {466, 4692}

\bibitem[\protect\citeauthoryear{{Kraljic} et~al.,}{{Kraljic} et~al.}{2018}]{Kraljic2018}
{Kraljic} K.,  et~al., 2018, \mn@doi [\mnras] {10.1093/mnras/stx2638}, \href {https://ui.adsabs.harvard.edu/abs/2018MNRAS.474..547K} {474, 547}

\bibitem[\protect\citeauthoryear{{Kreckel}, {Croxall}, {Groves}, {van de Weygaert}  \& {Pogge}}{{Kreckel} et~al.}{2015}]{Kreckel2015}
{Kreckel} K.,  {Croxall} K.,  {Groves} B.,  {van de Weygaert} R.,   {Pogge} R.~W.,  2015, \mn@doi [\apjl] {10.1088/2041-8205/798/1/L15}, \href {https://ui.adsabs.harvard.edu/abs/2015ApJ...798L..15K} {798, L15}

\bibitem[\protect\citeauthoryear{{Kroupa}}{{Kroupa}}{2001}]{Kroupa2001}
{Kroupa} P.,  2001, \mn@doi [\mnras] {10.1046/j.1365-8711.2001.04022.x}, \href {https://ui.adsabs.harvard.edu/abs/2001MNRAS.322..231K} {322, 231}

\bibitem[\protect\citeauthoryear{{Kuutma}, {Tamm}  \& {Tempel}}{{Kuutma} et~al.}{2017}]{Kuutma2017}
{Kuutma} T.,  {Tamm} A.,   {Tempel} E.,  2017, \mn@doi [\aap] {10.1051/0004-6361/201730526}, \href {https://ui.adsabs.harvard.edu/abs/2017A&A...600L...6K} {600, L6}

\bibitem[\protect\citeauthoryear{{Laigle} et~al.,}{{Laigle} et~al.}{2018}]{Laigle2018}
{Laigle} C.,  et~al., 2018, \mn@doi [\mnras] {10.1093/mnras/stx3055}, \href {https://ui.adsabs.harvard.edu/abs/2018MNRAS.474.5437L} {474, 5437}

\bibitem[\protect\citeauthoryear{Lee, Kashyap  \& Chu}{Lee et~al.}{1994}]{Lee1994}
Lee T.,  Kashyap R.,   Chu C.,  1994, \mn@doi [CVGIP: Graphical Models and Image Processing] {https://doi.org/10.1006/cgip.1994.1042}, 56, 462

\bibitem[\protect\citeauthoryear{{Lu}, {Mandelker}, {Oh}, {Dekel}, {van den Bosch}, {Springel}, {Nagai}  \& {van de Voort}}{{Lu} et~al.}{2024}]{LuY2024}
{Lu} Y.~S.,  {Mandelker} N.,  {Oh} S.~P.,  {Dekel} A.,  {van den Bosch} F.~C.,  {Springel} V.,  {Nagai} D.,   {van de Voort} F.,  2024, \mn@doi [\mnras] {10.1093/mnras/stad3779}, \href {https://ui.adsabs.harvard.edu/abs/2024MNRAS.52711256L} {527, 11256}

\bibitem[\protect\citeauthoryear{{Luber}, {van Gorkom}, {Hess}, {Pisano}, {Fern{\'a}ndez}  \& {Momjian}}{{Luber} et~al.}{2019}]{Luber2019}
{Luber} N.,  {van Gorkom} J.~H.,  {Hess} K.~M.,  {Pisano} D.~J.,  {Fern{\'a}ndez} X.,   {Momjian} E.,  2019, \mn@doi [\aj] {10.3847/1538-3881/ab1b6e}, \href {https://ui.adsabs.harvard.edu/abs/2019AJ....157..254L} {157, 254}

\bibitem[\protect\citeauthoryear{{Luo} et~al.,}{{Luo} et~al.}{2024}]{Luo2024}
{Luo} X.,  et~al., 2024, \mn@doi [\apj] {10.3847/1538-4357/ad392e}, \href {https://ui.adsabs.harvard.edu/abs/2024ApJ...966..236L} {966, 236}

\bibitem[\protect\citeauthoryear{{Malavasi} et~al.,}{{Malavasi} et~al.}{2017}]{Malavasi2017}
{Malavasi} N.,  et~al., 2017, \mn@doi [\mnras] {10.1093/mnras/stw2864}, \href {https://ui.adsabs.harvard.edu/abs/2017MNRAS.465.3817M} {465, 3817}

\bibitem[\protect\citeauthoryear{{Malavasi}, {Aghanim}, {Douspis}, {Tanimura}  \& {Bonjean}}{{Malavasi} et~al.}{2020}]{Malavasi2020}
{Malavasi} N.,  {Aghanim} N.,  {Douspis} M.,  {Tanimura} H.,   {Bonjean} V.,  2020, \mn@doi [\aap] {10.1051/0004-6361/202037647}, \href {https://ui.adsabs.harvard.edu/abs/2020A&A...642A..19M} {642, A19}

\bibitem[\protect\citeauthoryear{{Meert}, {Vikram}  \& {Bernardi}}{{Meert} et~al.}{2015}]{Meert2015}
{Meert} A.,  {Vikram} V.,   {Bernardi} M.,  2015, \mn@doi [\mnras] {10.1093/mnras/stu2333}, \href {https://ui.adsabs.harvard.edu/abs/2015MNRAS.446.3943M} {446, 3943}

\bibitem[\protect\citeauthoryear{{Montero-Dorta} \& {Rodriguez}}{{Montero-Dorta} \& {Rodriguez}}{2024}]{Montero2024}
{Montero-Dorta} A.~D.,  {Rodriguez} F.,  2024, \mn@doi [\mnras] {10.1093/mnras/stae796}, \href {https://ui.adsabs.harvard.edu/abs/2024MNRAS.531..290M} {531, 290}

\bibitem[\protect\citeauthoryear{{Naab}, {Johansson}  \& {Ostriker}}{{Naab} et~al.}{2009}]{Naab2009}
{Naab} T.,  {Johansson} P.~H.,   {Ostriker} J.~P.,  2009, \mn@doi [\apjl] {10.1088/0004-637X/699/2/L178}, \href {https://ui.adsabs.harvard.edu/abs/2009ApJ...699L.178N} {699, L178}

\bibitem[\protect\citeauthoryear{{Navarro}, {Frenk}  \& {White}}{{Navarro} et~al.}{1997}]{Navarro1997}
{Navarro} J.~F.,  {Frenk} C.~S.,   {White} S. D.~M.,  1997, \mn@doi [\apj] {10.1086/304888}, \href {https://ui.adsabs.harvard.edu/abs/1997ApJ...490..493N} {490, 493}

\bibitem[\protect\citeauthoryear{{O'Kane}, {Kuchner}, {Gray}  \& {Arag{\'o}n-Salamanca}}{{O'Kane} et~al.}{2024}]{Okane2024}
{O'Kane} C.~J.,  {Kuchner} U.,  {Gray} M.~E.,   {Arag{\'o}n-Salamanca} A.,  2024, \mn@doi [\mnras] {10.1093/mnras/stae2142}, \href {https://ui.adsabs.harvard.edu/abs/2024MNRAS.534.1682O} {534, 1682}

\bibitem[\protect\citeauthoryear{{Old} et~al.,}{{Old} et~al.}{2020}]{Old2020}
{Old} L.~J.,  et~al., 2020, \mn@doi [\mnras] {10.1093/mnras/staa579}, \href {https://ui.adsabs.harvard.edu/abs/2020MNRAS.493.5987O} {493, 5987}

\bibitem[\protect\citeauthoryear{{Pakmor} et~al.,}{{Pakmor} et~al.}{2023}]{Pakmor2023}
{Pakmor} R.,  et~al., 2023, \mn@doi [\mnras] {10.1093/mnras/stac3620}, \href {https://ui.adsabs.harvard.edu/abs/2023MNRAS.524.2539P} {524, 2539}

\bibitem[\protect\citeauthoryear{{Parente} et~al.,}{{Parente} et~al.}{2024}]{Parente2024}
{Parente} M.,  et~al., 2024, \mn@doi [\apj] {10.3847/1538-4357/ad320e}, \href {https://ui.adsabs.harvard.edu/abs/2024ApJ...966..154P} {966, 154}

\bibitem[\protect\citeauthoryear{{Pasha}, {Mandelker}, {van den Bosch}, {Springel}  \& {van de Voort}}{{Pasha} et~al.}{2023}]{Pasha2023}
{Pasha} I.,  {Mandelker} N.,  {van den Bosch} F.~C.,  {Springel} V.,   {van de Voort} F.,  2023, \mn@doi [\mnras] {10.1093/mnras/stac3776}, \href {https://ui.adsabs.harvard.edu/abs/2023MNRAS.520.2692P} {520, 2692}

\bibitem[\protect\citeauthoryear{{Peebles}}{{Peebles}}{1967}]{Peebles1967}
{Peebles} P.~J.~E.,  1967, \mn@doi [\apj] {10.1086/149077}, \href {https://ui.adsabs.harvard.edu/abs/1967ApJ...147..859P} {147, 859}

\bibitem[\protect\citeauthoryear{{Peebles}}{{Peebles}}{1969}]{Peebles1969}
{Peebles} P.~J.~E.,  1969, \mn@doi [\apj] {10.1086/149876}, \href {https://ui.adsabs.harvard.edu/abs/1969ApJ...155..393P} {155, 393}

\bibitem[\protect\citeauthoryear{{Pfeifer}, {Libeskind}, {Hoffman}, {Hellwing}, {Bilicki}  \& {Naidoo}}{{Pfeifer} et~al.}{2022}]{Pfeifer2022}
{Pfeifer} S.,  {Libeskind} N.~I.,  {Hoffman} Y.,  {Hellwing} W.~A.,  {Bilicki} M.,   {Naidoo} K.,  2022, \mn@doi [\mnras] {10.1093/mnras/stac1382}, \href {https://ui.adsabs.harvard.edu/abs/2022MNRAS.514..470P} {514, 470}

\bibitem[\protect\citeauthoryear{{Rosas-Guevara}, {Tissera}, {Lagos}, {Paillas}  \& {Padilla}}{{Rosas-Guevara} et~al.}{2022}]{Rosas2022}
{Rosas-Guevara} Y.,  {Tissera} P.,  {Lagos} C. d.~P.,  {Paillas} E.,   {Padilla} N.,  2022, \mn@doi [\mnras] {10.1093/mnras/stac2583}, \href {https://ui.adsabs.harvard.edu/abs/2022MNRAS.517..712R} {517, 712}

\bibitem[\protect\citeauthoryear{{Rost}, {Stasyszyn}, {Pereyra}  \& {Mart{\'\i}nez}}{{Rost} et~al.}{2020}]{Rost2020}
{Rost} A.,  {Stasyszyn} F.,  {Pereyra} L.,   {Mart{\'\i}nez} H.~J.,  2020, \mn@doi [\mnras] {10.1093/mnras/staa320}, \href {https://ui.adsabs.harvard.edu/abs/2020MNRAS.493.1936R} {493, 1936}

\bibitem[\protect\citeauthoryear{{Rowntree} et~al.,}{{Rowntree} et~al.}{2024}]{Rowntree2024}
{Rowntree} A.~R.,  et~al., 2024, \mn@doi [\mnras] {10.1093/mnras/stae1384}, \href {https://ui.adsabs.harvard.edu/abs/2024MNRAS.531.3858R} {531, 3858}

\bibitem[\protect\citeauthoryear{{Salerno} et~al.,}{{Salerno} et~al.}{2020}]{Salerno2020}
{Salerno} J.~M.,  et~al., 2020, \mn@doi [\mnras] {10.1093/mnras/staa545}, \href {https://ui.adsabs.harvard.edu/abs/2020MNRAS.493.4950S} {493, 4950}

\bibitem[\protect\citeauthoryear{{Sarron}, {Adami}, {Durret}  \& {Laigle}}{{Sarron} et~al.}{2019}]{Sarron2019}
{Sarron} F.,  {Adami} C.,  {Durret} F.,   {Laigle} C.,  2019, \mn@doi [\aap] {10.1051/0004-6361/201935394}, \href {https://ui.adsabs.harvard.edu/abs/2019A&A...632A..49S} {632, A49}

\bibitem[\protect\citeauthoryear{{Simard}, {Mendel}, {Patton}, {Ellison}  \& {McConnachie}}{{Simard} et~al.}{2011}]{Simard2011}
{Simard} L.,  {Mendel} J.~T.,  {Patton} D.~R.,  {Ellison} S.~L.,   {McConnachie} A.~W.,  2011, \mn@doi [\apjs] {10.1088/0067-0049/196/1/11}, \href {https://ui.adsabs.harvard.edu/abs/2011ApJS..196...11S} {196, 11}

\bibitem[\protect\citeauthoryear{{Sousbie}}{{Sousbie}}{2011}]{Sousbie2011a}
{Sousbie} T.,  2011, \mn@doi [\mnras] {10.1111/j.1365-2966.2011.18394.x}, \href {https://ui.adsabs.harvard.edu/abs/2011MNRAS.414..350S} {414, 350}

\bibitem[\protect\citeauthoryear{{Sousbie}, {Pichon}  \& {Kawahara}}{{Sousbie} et~al.}{2011}]{Sousbie2011b}
{Sousbie} T.,  {Pichon} C.,   {Kawahara} H.,  2011, \mn@doi [\mnras] {10.1111/j.1365-2966.2011.18395.x}, \href {https://ui.adsabs.harvard.edu/abs/2011MNRAS.414..384S} {414, 384}

\bibitem[\protect\citeauthoryear{{Springel}}{{Springel}}{2005}]{Springel2005}
{Springel} V.,  2005, \mn@doi [\mnras] {10.1111/j.1365-2966.2005.09655.x}, \href {https://ui.adsabs.harvard.edu/abs/2005MNRAS.364.1105S} {364, 1105}

\bibitem[\protect\citeauthoryear{{Springel}, {White}, {Tormen}  \& {Kauffmann}}{{Springel} et~al.}{2001}]{Springel2001}
{Springel} V.,  {White} S. D.~M.,  {Tormen} G.,   {Kauffmann} G.,  2001, \mn@doi [\mnras] {10.1046/j.1365-8711.2001.04912.x}, \href {https://ui.adsabs.harvard.edu/abs/2001MNRAS.328..726S} {328, 726}

\bibitem[\protect\citeauthoryear{{Taamoli} et~al.,}{{Taamoli} et~al.}{2024}]{Taamoli2024}
{Taamoli} S.,  et~al., 2024, \mn@doi [\apj] {10.3847/1538-4357/ad32c5}, \href {https://ui.adsabs.harvard.edu/abs/2024ApJ...966...18T} {966, 18}

\bibitem[\protect\citeauthoryear{{Tugay} \& {Tarnopolski}}{{Tugay} \& {Tarnopolski}}{2023}]{Tugay2023}
{Tugay} A.,  {Tarnopolski} M.,  2023, \mn@doi [\apj] {10.3847/1538-4357/acd9a4}, \href {https://ui.adsabs.harvard.edu/abs/2023ApJ...952....3T} {952, 3}

\bibitem[\protect\citeauthoryear{{Tweed}, {Yang}, {Wang}, {Cui}, {Zhang}, {Li}, {Jing}  \& {Mo}}{{Tweed} et~al.}{2017}]{Tweed2017}
{Tweed} D.,  {Yang} X.,  {Wang} H.,  {Cui} W.,  {Zhang} Y.,  {Li} S.,  {Jing} Y.~P.,   {Mo} H.~J.,  2017, \mn@doi [\apj] {10.3847/1538-4357/aa6bf8}, \href {https://ui.adsabs.harvard.edu/abs/2017ApJ...841...55T} {841, 55}

\bibitem[\protect\citeauthoryear{{Vulcani} et~al.,}{{Vulcani} et~al.}{2019}]{Vulcani2019}
{Vulcani} B.,  et~al., 2019, \mn@doi [\mnras] {10.1093/mnras/stz1399}, \href {https://ui.adsabs.harvard.edu/abs/2019MNRAS.487.2278V} {487, 2278}

\bibitem[\protect\citeauthoryear{{Wang}, {Mo}, {Yang}  \& {van den Bosch}}{{Wang} et~al.}{2012}]{WangHuiyuan2012}
{Wang} H.,  {Mo} H.~J.,  {Yang} X.,   {van den Bosch} F.~C.,  2012, \mn@doi [\mnras] {10.1111/j.1365-2966.2011.20174.x}, \href {https://ui.adsabs.harvard.edu/abs/2012MNRAS.420.1809W} {420, 1809}

\bibitem[\protect\citeauthoryear{{Wang}, {Mo}, {Yang}, {Jing}  \& {Lin}}{{Wang} et~al.}{2014}]{WangHuiyuan2014}
{Wang} H.,  {Mo} H.~J.,  {Yang} X.,  {Jing} Y.~P.,   {Lin} W.~P.,  2014, \mn@doi [\apj] {10.1088/0004-637X/794/1/94}, \href {https://ui.adsabs.harvard.edu/abs/2014ApJ...794...94W} {794, 94}

\bibitem[\protect\citeauthoryear{{Wang} et~al.,}{{Wang} et~al.}{2016}]{WangHuiyuan2016}
{Wang} H.,  et~al., 2016, \mn@doi [\apj] {10.3847/0004-637X/831/2/164}, \href {https://ui.adsabs.harvard.edu/abs/2016ApJ...831..164W} {831, 164}

\bibitem[\protect\citeauthoryear{{Wang} et~al.,}{{Wang} et~al.}{2018}]{WangHY2018}
{Wang} H.,  et~al., 2018, \mn@doi [\apj] {10.3847/1538-4357/aa9e01}, \href {https://ui.adsabs.harvard.edu/abs/2018ApJ...852...31W} {852, 31}

\bibitem[\protect\citeauthoryear{{Wang}, {Shi}, {Yang}, {Li}, {Liu}  \& {Li}}{{Wang} et~al.}{2024a}]{WangZitong2024}
{Wang} Z.,  {Shi} F.,  {Yang} X.,  {Li} Q.,  {Liu} Y.,   {Li} X.,  2024a, \mn@doi [Science China Physics, Mechanics, and Astronomy] {10.1007/s11433-023-2192-9}, \href {https://ui.adsabs.harvard.edu/abs/2024SCPMA..6719513W} {67, 219513}

\bibitem[\protect\citeauthoryear{{Wang} et~al.,}{{Wang} et~al.}{2024b}]{WangWei2024}
{Wang} W.,  et~al., 2024b, \mn@doi [\mnras] {10.1093/mnras/stae1801}, \href {https://ui.adsabs.harvard.edu/abs/2024MNRAS.532.4604W} {532, 4604}

\bibitem[\protect\citeauthoryear{{Wechsler} \& {Tinker}}{{Wechsler} \& {Tinker}}{2018}]{Wechsler2018}
{Wechsler} R.~H.,  {Tinker} J.~L.,  2018, \mn@doi [\araa] {10.1146/annurev-astro-081817-051756}, \href {https://ui.adsabs.harvard.edu/abs/2018ARA&A..56..435W} {56, 435}

\bibitem[\protect\citeauthoryear{{Wegner}, {Salzer}, {Taylor}  \& {Hirschauer}}{{Wegner} et~al.}{2019}]{Wegner2019}
{Wegner} G.~A.,  {Salzer} J.~J.,  {Taylor} J.~M.,   {Hirschauer} A.~S.,  2019, \mn@doi [\apj] {10.3847/1538-4357/ab3a3c}, \href {https://ui.adsabs.harvard.edu/abs/2019ApJ...883...29W} {883, 29}

\bibitem[\protect\citeauthoryear{{Willett} et~al.,}{{Willett} et~al.}{2013}]{Willett2013}
{Willett} K.~W.,  et~al., 2013, \mn@doi [\mnras] {10.1093/mnras/stt1458}, \href {https://ui.adsabs.harvard.edu/abs/2013MNRAS.435.2835W} {435, 2835}

\bibitem[\protect\citeauthoryear{{Winkel}, {Pasquali}, {Kraljic}, {Smith}, {Gallazzi}  \& {Jackson}}{{Winkel} et~al.}{2021}]{Winkel2021}
{Winkel} N.,  {Pasquali} A.,  {Kraljic} K.,  {Smith} R.,  {Gallazzi} A.,   {Jackson} T.~M.,  2021, \mn@doi [\mnras] {10.1093/mnras/stab1562}, \href {https://ui.adsabs.harvard.edu/abs/2021MNRAS.505.4920W} {505, 4920}

\bibitem[\protect\citeauthoryear{{Woo} et~al.,}{{Woo} et~al.}{2013}]{Woo2013}
{Woo} J.,  et~al., 2013, \mn@doi [\mnras] {10.1093/mnras/sts274}, \href {https://ui.adsabs.harvard.edu/abs/2013MNRAS.428.3306W} {428, 3306}

\bibitem[\protect\citeauthoryear{{Wright} et~al.,}{{Wright} et~al.}{2010}]{Wright2010}
{Wright} E.~L.,  et~al., 2010, \mn@doi [\aj] {10.1088/0004-6256/140/6/1868}, \href {https://ui.adsabs.harvard.edu/abs/2010AJ....140.1868W} {140, 1868}

\bibitem[\protect\citeauthoryear{{Xue}, {Rong}  \& {Hua}}{{Xue} et~al.}{2024}]{XueWX2024}
{Xue} W.,  {Rong} Y.,   {Hua} Z.,  2024, \mn@doi [arXiv e-prints] {10.48550/arXiv.2411.11443}, \href {https://ui.adsabs.harvard.edu/abs/2024arXiv241111443X} {p. arXiv:2411.11443}

\bibitem[\protect\citeauthoryear{{Yang}, {Mo}, {van den Bosch}, {Pasquali}, {Li}  \& {Barden}}{{Yang} et~al.}{2007}]{Yang2007}
{Yang} X.,  {Mo} H.~J.,  {van den Bosch} F.~C.,  {Pasquali} A.,  {Li} C.,   {Barden} M.,  2007, \mn@doi [\apj] {10.1086/522027}, \href {https://ui.adsabs.harvard.edu/abs/2007ApJ...671..153Y} {671, 153}

\bibitem[\protect\citeauthoryear{{Yang}, {Mo}, {van den Bosch}, {Zhang}  \& {Han}}{{Yang} et~al.}{2012}]{Yang2012}
{Yang} X.,  {Mo} H.~J.,  {van den Bosch} F.~C.,  {Zhang} Y.,   {Han} J.,  2012, \mn@doi [\apj] {10.1088/0004-637X/752/1/41}, \href {https://ui.adsabs.harvard.edu/abs/2012ApJ...752...41Y} {752, 41}

\bibitem[\protect\citeauthoryear{{Yang} et~al.,}{{Yang} et~al.}{2018}]{Yang2018}
{Yang} X.,  et~al., 2018, \mn@doi [\apj] {10.3847/1538-4357/aac2ce}, \href {https://ui.adsabs.harvard.edu/abs/2018ApJ...860...30Y} {860, 30}

\bibitem[\protect\citeauthoryear{{Yang}, {Hudson}  \& {Afshordi}}{{Yang} et~al.}{2022}]{YangTianyi2022}
{Yang} T.,  {Hudson} M.~J.,   {Afshordi} N.,  2022, \mn@doi [\mnras] {10.1093/mnras/stac2564}, \href {https://ui.adsabs.harvard.edu/abs/2022MNRAS.516.6041Y} {516, 6041}

\bibitem[\protect\citeauthoryear{{York} et~al.,}{{York} et~al.}{2000}]{York2000}
{York} D.~G.,  et~al., 2000, \mn@doi [\aj] {10.1086/301513}, \href {https://ui.adsabs.harvard.edu/abs/2000AJ....120.1579Y} {120, 1579}

\bibitem[\protect\citeauthoryear{{Zakharova}, {Vulcani}, {De Lucia}, {Xie}, {Hirschmann}  \& {Fontanot}}{{Zakharova} et~al.}{2023}]{Zakharova2023}
{Zakharova} D.,  {Vulcani} B.,  {De Lucia} G.,  {Xie} L.,  {Hirschmann} M.,   {Fontanot} F.,  2023, \mn@doi [\mnras] {10.1093/mnras/stad2562}, \href {https://ui.adsabs.harvard.edu/abs/2023MNRAS.525.4079Z} {525, 4079}

\bibitem[\protect\citeauthoryear{Zhang \& Suen}{Zhang \& Suen}{1984}]{Zhang1984}
Zhang T.~Y.,  Suen C.~Y.,  1984, \mn@doi [Commun. ACM] {10.1145/357994.358023}, 27, 236–239

\bibitem[\protect\citeauthoryear{{Zhang} \& {Yang}}{{Zhang} \& {Yang}}{2019}]{Zhang2019}
{Zhang} Y.-C.,  {Yang} X.-H.,  2019, \mn@doi [Research in Astronomy and Astrophysics] {10.1088/1674-4527/19/1/6}, \href {https://ui.adsabs.harvard.edu/abs/2019RAA....19....6Z} {19, 006}

\bibitem[\protect\citeauthoryear{{Zhang}, {Yang}, {Faltenbacher}, {Springel}, {Lin}  \& {Wang}}{{Zhang} et~al.}{2009}]{Zhang2009}
{Zhang} Y.,  {Yang} X.,  {Faltenbacher} A.,  {Springel} V.,  {Lin} W.,   {Wang} H.,  2009, \mn@doi [\apj] {10.1088/0004-637X/706/1/747}, \href {https://ui.adsabs.harvard.edu/abs/2009ApJ...706..747Z} {706, 747}

\bibitem[\protect\citeauthoryear{{Zhang}, {Yang}, {Wang}, {Wang}, {Mo}  \& {van den Bosch}}{{Zhang} et~al.}{2013}]{Zhang2013}
{Zhang} Y.,  {Yang} X.,  {Wang} H.,  {Wang} L.,  {Mo} H.~J.,   {van den Bosch} F.~C.,  2013, \mn@doi [\apj] {10.1088/0004-637X/779/2/160}, \href {https://ui.adsabs.harvard.edu/abs/2013ApJ...779..160Z} {779, 160}

\bibitem[\protect\citeauthoryear{{Zhang}, {Yang}, {Wang}, {Wang}, {Luo}, {Mo}  \& {van den Bosch}}{{Zhang} et~al.}{2015}]{Zhang2015}
{Zhang} Y.,  {Yang} X.,  {Wang} H.,  {Wang} L.,  {Luo} W.,  {Mo} H.~J.,   {van den Bosch} F.~C.,  2015, \mn@doi [\apj] {10.1088/0004-637X/798/1/17}, \href {https://ui.adsabs.harvard.edu/abs/2015ApJ...798...17Z} {798, 17}

\bibitem[\protect\citeauthoryear{{Zhang}, {Yang}  \& {Guo}}{{Zhang} et~al.}{2021a}]{Zhang2021a}
{Zhang} Y.,  {Yang} X.,   {Guo} H.,  2021a, \mn@doi [\mnras] {10.1093/mnras/staa2356}, \href {https://ui.adsabs.harvard.edu/abs/2021MNRAS.500.1895Z} {500, 1895}

\bibitem[\protect\citeauthoryear{{Zhang}, {Yang}  \& {Guo}}{{Zhang} et~al.}{2021b}]{Zhang2021b}
{Zhang} Y.,  {Yang} X.,   {Guo} H.,  2021b, \mn@doi [\mnras] {10.1093/mnras/stab2487}, \href {https://ui.adsabs.harvard.edu/abs/2021MNRAS.507.5320Z} {507, 5320}

\bibitem[\protect\citeauthoryear{{Zhang}, {Yang}  \& {Guo}}{{Zhang} et~al.}{2022}]{Zhang2022}
{Zhang} Y.,  {Yang} X.,   {Guo} H.,  2022, \mn@doi [\mnras] {10.1093/mnras/stac2934}, \href {https://ui.adsabs.harvard.edu/abs/2022MNRAS.517.3579Z} {517, 3579}

\bibitem[\protect\citeauthoryear{{Zhang}, {Guo}, {Yang}  \& {Wang}}{{Zhang} et~al.}{2024}]{Zhang2024}
{Zhang} Y.,  {Guo} H.,  {Yang} X.,   {Wang} P.,  2024, \mn@doi [\mnras] {10.1093/mnras/stae1914}, \href {https://ui.adsabs.harvard.edu/abs/2024MNRAS.533.1048Z} {533, 1048}

\bibitem[\protect\citeauthoryear{{Zhao}, {Jing}, {Mo}  \& {B{\"o}rner}}{{Zhao} et~al.}{2009}]{Zhao2009}
{Zhao} D.~H.,  {Jing} Y.~P.,  {Mo} H.~J.,   {B{\"o}rner} G.,  2009, \mn@doi [\apj] {10.1088/0004-637X/707/1/354}, \href {https://ui.adsabs.harvard.edu/abs/2009ApJ...707..354Z} {707, 354}

\bibitem[\protect\citeauthoryear{{da Cunha}, {Charlot}  \& {Elbaz}}{{da Cunha} et~al.}{2008}]{daCunha2008}
{da Cunha} E.,  {Charlot} S.,   {Elbaz} D.,  2008, \mn@doi [\mnras] {10.1111/j.1365-2966.2008.13535.x}, \href {https://ui.adsabs.harvard.edu/abs/2008MNRAS.388.1595D} {388, 1595}

\makeatother
\end{thebibliography}

\bsp    
\label{lastpage}
\end{document}